\documentclass[aip,apl,reprint,superscriptaddress,floatfix]{revtex4-2}

\usepackage{amsmath,amssymb}
\usepackage{graphicx}
\usepackage{bm}
\usepackage{siunitx}
\usepackage{xcolor}
\usepackage{url}

\begin{document}

\title{Exact branch-transfer criterion for common-mode Thomson heat cancellation in thermoelectric couples}

\author{Peng Kang}
\affiliation{School of Materials Science and Engineering, Beihang University,
 No.~37 Xueyuan Road, Beijing 100191, China}
\affiliation{National Key Laboratory of Artificial Intelligence for Material Science,
 Beihang University, No.~37 Xueyuan Road, Beijing 100191, China}
\affiliation{Tianmushan Laboratory, Beihang University, Hangzhou 311115, China}

\author{Da Wan}
\affiliation{School of Materials Science and Engineering, Beihang University,
 No.~37 Xueyuan Road, Beijing 100191, China}
\affiliation{National Key Laboratory of Artificial Intelligence for Material Science,
 Beihang University, No.~37 Xueyuan Road, Beijing 100191, China}
\affiliation{Tianmushan Laboratory, Beihang University, Hangzhou 311115, China}

\author{Shulin Bai}
\affiliation{School of Materials Science and Engineering, Beihang University,
 No.~37 Xueyuan Road, Beijing 100191, China}
\affiliation{National Key Laboratory of Artificial Intelligence for Material Science,
 Beihang University, No.~37 Xueyuan Road, Beijing 100191, China}
\affiliation{Tianmushan Laboratory, Beihang University, Hangzhou 311115, China}
\affiliation{Center for Bioinspired Science and Technology, Hangzhou International
 Innovation Institute, Beihang University, Hangzhou 311115, China}

\author{Wei Yin}
\affiliation{Tianmushan Laboratory, Beihang University, Hangzhou 311115, China}

\author{Peng Wang}
\affiliation{School of Materials Science and Engineering, Beihang University,
 No.~37 Xueyuan Road, Beijing 100191, China}
\affiliation{National Key Laboratory of Artificial Intelligence for Material Science,
 Beihang University, No.~37 Xueyuan Road, Beijing 100191, China}

\author{Chenglong Wen}
\affiliation{School of Materials Science and Engineering, Beihang University,
 No.~37 Xueyuan Road, Beijing 100191, China}
\affiliation{National Key Laboratory of Artificial Intelligence for Material Science,
 Beihang University, No.~37 Xueyuan Road, Beijing 100191, China}

\author{Zhen Li}
\affiliation{School of Materials Science and Engineering, Beihang University,
 No.~37 Xueyuan Road, Beijing 100191, China}
\affiliation{National Key Laboratory of Artificial Intelligence for Material Science,
 Beihang University, No.~37 Xueyuan Road, Beijing 100191, China}
\affiliation{Tianmushan Laboratory, Beihang University, Hangzhou 311115, China}

\author{Yu Liu}
\affiliation{School of Materials Science and Engineering, Beihang University,
 No.~37 Xueyuan Road, Beijing 100191, China}
\affiliation{National Key Laboratory of Artificial Intelligence for Material Science,
 Beihang University, No.~37 Xueyuan Road, Beijing 100191, China}
\affiliation{Tianmushan Laboratory, Beihang University, Hangzhou 311115, China}

\author{Lei Zheng}
\email{zhenglei@buaa.edu.cn}
\affiliation{School of Materials Science and Engineering, Beihang University,
 No.~37 Xueyuan Road, Beijing 100191, China}
\affiliation{National Key Laboratory of Artificial Intelligence for Material Science,
 Beihang University, No.~37 Xueyuan Road, Beijing 100191, China}
\affiliation{Tianmushan Laboratory, Beihang University, Hangzhou 311115, China}

\author{Li-Dong Zhao}
\email{zhaolidong@buaa.edu.cn}
\affiliation{School of Materials Science and Engineering, Beihang University,
 No.~37 Xueyuan Road, Beijing 100191, China}
\affiliation{National Key Laboratory of Artificial Intelligence for Material Science,
 Beihang University, No.~37 Xueyuan Road, Beijing 100191, China}
\affiliation{Tianmushan Laboratory, Beihang University, Hangzhou 311115, China}
\affiliation{Center for Bioinspired Science and Technology, Hangzhou International
 Innovation Institute, Beihang University, Hangzhou 311115, China}

\date{}

\begin{abstract}
Thermoelectric p- and n-type legs are commonly paired by matching their Seebeck
magnitudes, although a cooler responds to heat transported through its complete
electrical and thermal network.  We decompose the leg coefficients into
differential thermopower $\alpha=S_p-S_n$ and common thermopower
$M=(S_p+S_n)/2$.  In a connected steady-state scalar thermoelectric network, a temperature-independent
co-shift applied to every electrically active segment is an exact terminal null.
A temperature-dependent perturbation of the legs relative to fixed leads is instead
physical.  At fixed current and shared isothermal endpoints, its first-order
cold-port response is the action of $\Gamma_m=T\,dm/dT$ on the difference between
the p- and n-branch oriented collection measures.  We prove that every continuous
$\Gamma_m$ cancels if and only if these measures are equal.  In the constant-property,
linear-common-mode limit, matching $R_i/K_i^{\rm leg}$ is sufficient and does not require identical
legs.  One- and two-dimensional calculations confirm the analytic reductions
within their stated domains.  For split thermal pads, the analysis gives the
exact array law
$\Delta Q_{c,\Sigma}=\sum_j C_jI_j\Delta T_{c,j}$ and, for series elements with
isothermal hot pairs, $I\Delta V_\Sigma=-\Delta Q_{c,\Sigma}$.  A representative
seven-pair model gives corresponding increments of 7.87 mW and $-2.80$ mV.  Branch
transfer and endpoint topology therefore provide distinct material-pairing and
device-test criteria for common-mode Thomson heat.
\end{abstract}

\maketitle

Recent thermoelectric materials research spans electronic-structure design,
strain-controlled lattice dynamics, higher-order phonon scattering, moir\'e control,
and data-guided screening.\cite{bai2026designing,wan2025machine,wan2025strain,
bai2025rashba,kang2026twisted} These strategies reshape the coupled temperature-
dependent transport functions.  Their device benefit, however, is determined by
how a p--n pair transfers heat to the cold port.
\cite{bell2008systems,kim2017bridge} Matching the leg Seebeck
magnitudes is a useful heuristic within related material families,
\cite{barabash2014matching} but cooling also depends on electrical resistivity,
thermal conductivity, geometry, current, interfaces, and network topology.
\cite{snyder2003compatibility,pan2021pairing,zhou2023structure}

Absolute-thermopower measurements fix the reference scale, while the Kelvin
relations connect Seebeck gradients to Thomson heat.
\cite{callen1948onsager,domenicali1954irreversible,roberts1977absolute,
amagai2019absolute,apertet2016kelvin}
Temperature-dependent cooling models describe Thomson heat redistribution within
individual elements, while thermoelectric adjoint methods evaluate the sensitivity
of a prescribed device objective.
\cite{snyder2012thomson,ryu2021three,zhao2026nonreciprocal,
lundgaard2018adjoint,reales2024adjoint} Network formulations also resolve spatial
inhomogeneity and measurement-protocol bias.\cite{angst2016network} These
approaches do not give a paired-branch cancellation
criterion in temperature space.  The unresolved question is when a common p/n
Seebeck gradient cancels between the branches and when it survives at the cold
port.  Answering it requires a clear distinction between a shift of the complete
electrical network and a change of the thermoelectric legs relative to fixed leads
and interconnects.

Here we derive an oriented cold-port collection measure for each branch.  Equality
of the p- and n-branch measures is necessary and sufficient for cancellation of
every continuous common-mode Thomson perturbation.  A constant-property reduction
then gives a transfer-matching rule that is weaker than leg identity.  We connect
the theorem to an exact split-pad heat--voltage relation and assess its range of
validity using
nonlinear one-dimensional, finite-contact, and heterogeneous two-dimensional
calculations.  Applications to published transport data and the corresponding
data requirements are reported in the supplementary material.

\begin{figure*}[t]
 \centering
 \includegraphics[width=6.69in]{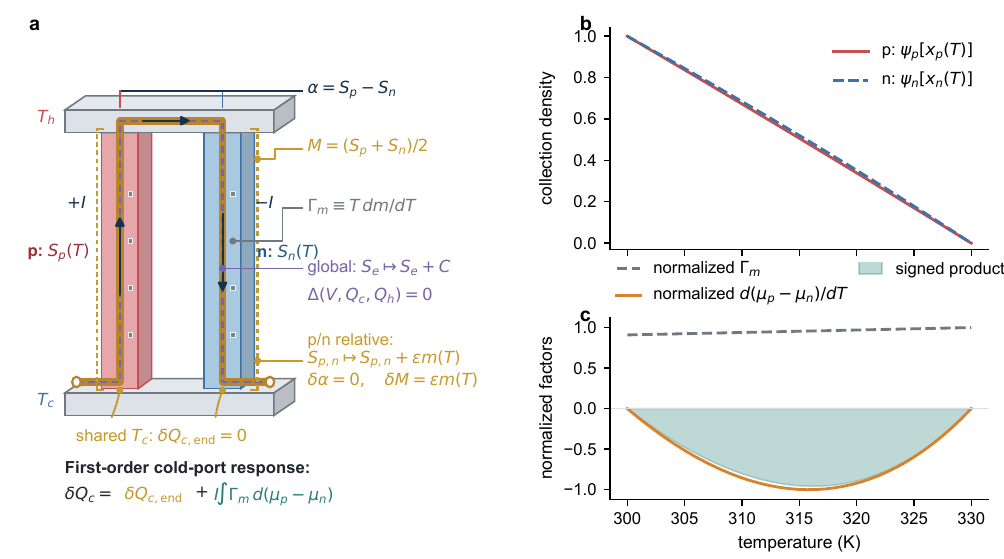}
 \caption{\label{fig:collection}\textbf{Common-mode Thomson heat is collected through
 branch-transfer mismatch.} (a) An axonometric p--n couple is electrically in series
 and thermally in parallel between shared isothermal ports.  The complete active
 electrical network is shifted in the global-null operation, whereas the branch-
 relative operation selects only the p/n legs against fixed references.  The arrows
 denote $I_p=+I$ and $I_n=-I$; matching cold temperatures remove the direct endpoint
 term. (b) Oriented cold-port collection densities. (c) Normalized
 $\Gamma_m(T)$, the collection-density difference, and their signed overlap.  For
 the constant-property reference calculation specified in the supplementary
 material, the
 two branch transfers are $+2.347$ and $-2.391$ mW, leaving $-43.8~\mu$W per
 unit perturbation.}
\end{figure*}

We write
\begin{equation}
 \begin{aligned}
  \alpha(T)&=S_p(T)-S_n(T), &
  M(T)&=\frac{S_p(T)+S_n(T)}{2},\\
  \Gamma_{\rm cm}(T)&=T\frac{dM}{dT}. &&
 \end{aligned}
 \label{eq:coordinates_strengthened}
\end{equation}
Two transformations must be distinguished (Fig.~\ref{fig:collection}a).  For a
steady branch with area $A$, signed current $I$, electric potential $\phi$, heat rate
$H$, and cold-to-hot coordinate $x$, the local transport laws are
\begin{equation}
 \begin{aligned}
  \phi'&=-\rho I/A-ST', & H&=STI-\kappa AT',\\
  H'&=I(-\phi'). &&
 \end{aligned}
 \label{eq:local_strengthened}
\end{equation}
Under the global constant co-shift $S^*=S+C$, define the transformed fields
\begin{equation}
 \phi^*=\phi-C(T-T_{\rm ref}),\qquad H^*=H+CTI,
 \label{eq:global_shift}
\end{equation}
which preserve Eq.~(\ref{eq:local_strengthened}) pointwise.  When the shift includes every
electrically active segment, including leads and interconnects,
$\sum_j\Delta H_j=CT\sum_jI_j=0$ at each node and the voltage shift is an exact
differential.  The complete terminal $I$--$V$--$Q$ map is therefore unchanged.

A different operation changes the p/n branches relative to fixed leads or
interconnects.  Let $S_i^\epsilon=S_i+\epsilon m(T)$ at fixed current and fixed branch
endpoint temperatures, and define the perturbation profile
$\Gamma_m(T)=T\,dm/dT$.  The Thomson coefficient then changes by
$\epsilon\Gamma_m(T)$ and $T_i^\epsilon=T_i+\epsilon y_i+O(\epsilon^2)$.  We consider
a nondegenerate baseline for which the homogeneous Dirichlet linearization has a
trivial kernel; an explicit sufficient condition is given in the supplementary
material, and every one-dimensional state used here satisfies it.  Linearization
gives $\mathcal L_i y_i=I_i\Gamma_m(T_i)T_i'$.  The cold-port
adjoint field satisfies
\begin{equation}
 \begin{aligned}
  \mathcal L_i^\dagger\psi_i
  &=K_i(T_i)\psi_i''+I_i\tau_i(T_i)\psi_i'
    +r_{i,T}(T_i)I_i^2\psi_i=0,\\
  \psi_i(0)&=1,\qquad \psi_i(L_i)=0,
 \end{aligned}
 \label{eq:adjoint}
\end{equation}
where $K_i=\kappa_iA_i$, $r_i=\rho_i/A_i$, and
$\tau_i=T\,dS_i/dT$.  Define the oriented temperature-space collection measure
\begin{equation}
 \mu_i(B)=\int_{\{x:T_i(x)\in B\}}\psi_i(x)T_i'(x)\,dx.
 \label{eq:measure}
\end{equation}
The first-order aggregate cold-side response is then
\begin{equation}
 \boxed{\delta Q_c=\delta Q_{c,\mathrm{end}}
 +I\int\Gamma_m(T)\,d(\mu_p-\mu_n)},
 \label{eq:kernel}
\end{equation}
with
$\delta Q_{c,\mathrm{end}}=I[T_{c,p}m(T_{c,p})-T_{c,n}m(T_{c,n})]$.
For matching endpoint temperatures, the endpoint term vanishes.  A specified
$\Gamma_m$ cancels when it is orthogonal to $\mu_p-\mu_n$.  Cancellation holds for
every continuous $\Gamma_m$ if and only if $\mu_p=\mu_n$.  Single-element
Thomson models determine heat redistribution for a specified element,
\cite{snyder2012thomson,zhao2026nonreciprocal} whereas thermoelectric adjoints
evaluate the sensitivity of a prescribed device objective.
\cite{lundgaard2018adjoint,reales2024adjoint} Neither result implies equality of
the two push-forward measures required for profile-independent p/n cancellation;
their difference is the new paired-branch object.  Equality of the
oriented baseline and adjoint problems is sufficient, but not necessary, for
measure equality.  The oriented form also covers nonmonotonic $T_i(x)$, for which
a single-valued inverse $x_i(T)$ does not exist.

The theorem also yields a physically transparent device-scale reduction.  For
constant $\rho_i$ and $\kappa_i$,
fixed endpoints, and a constant-Seebeck baseline, let $u=x/L_i$ and
\begin{equation}
 \begin{aligned}
  g_i&=\frac{\rho_iI^2L_i^2}{2\kappa_iA_i^2}, &
  T_i(u)&=T_c+\Delta T\,u+g_i u(1-u),\\
  \psi_i&=1-u. &&
 \end{aligned}
\end{equation}
For $m(T)=b(T-T_c)$, direct integration gives
\begin{equation}
 \delta Q_c=Ib\,(g_p-g_n)
 \left(\frac{T_c}{6}+\frac{\Delta T}{12}
 +\frac{g_p+g_n}{60}\right).
 \label{eq:cubic}
\end{equation}
The leading response is therefore proportional to
$I^3b[R_p/K_p^{\rm leg}-R_n/K_n^{\rm leg}]$, while the last term supplies the
exact finite-current $I^5$ correction.  Since
\begin{equation}
 g_i=\frac{I^2}{2}\frac{R_i}{K_i^{\rm leg}},\qquad
 R_i=\frac{\rho_iL_i}{A_i},\quad K_i^{\rm leg}=\frac{\kappa_iA_i}{L_i},
 \label{eq:transfer_matching}
\end{equation}
nonidentical legs with matched $R_i/K_i^{\rm leg}$ have $g_p=g_n$ and cancel
the entire linear-$m(T)$ family in this limit.  Thus branch identity is not the
relevant condition; the required match is between electrical heating and thermal
transfer.  Analytic, adjoint, nonlinear one-dimensional, and finite-contact checks
of this reduction are given in the supplementary material.

The same branch-relative transformation yields an exact boundary law when the
cold endpoints are split.  Consider p/n endpoint pads whose thermopowers are
changed relative to calibrated fixed reference leads.  For an array with
element-dependent constant contrast $C_j$, local current $I_j$, and p--n endpoint
splits $\Delta T_{c,j}$ and $\Delta T_{h,j}$,
\begin{align}
 \Delta Q_{c,\Sigma}&=\sum_j C_jI_j\Delta T_{c,j},\nonumber\\
 \Delta Q_{h,\Sigma}&=\sum_j C_jI_j\Delta T_{h,j},\nonumber\\
 \Delta P_{\mathrm{in},\Sigma}
 &=\sum_j C_jI_j(\Delta T_{h,j}-\Delta T_{c,j}).
 \label{eq:split_pad}
\end{align}
These finite increments obey
$\Delta Q_{h,\Sigma}-\Delta Q_{c,\Sigma}-\Delta P_{\mathrm{in},\Sigma}=0$
exactly.  Here $Q_{c,\Sigma}$ is the sum of the heat rates at two separately
controlled cold pads, not the heat extracted from a single isothermal reservoir.
For series elements
with $I_j=I$ and $\Delta T_{h,j}=0$, let
$\Delta V_\Sigma=\sum_j\Delta V_j$ denote the total series-terminal increment.
The same transformation then gives the directly testable identity
$I\Delta V_\Sigma=-\Delta Q_{c,\Sigma}$.

\begin{figure}[!t]
 \centering
 \includegraphics[width=0.97\columnwidth]{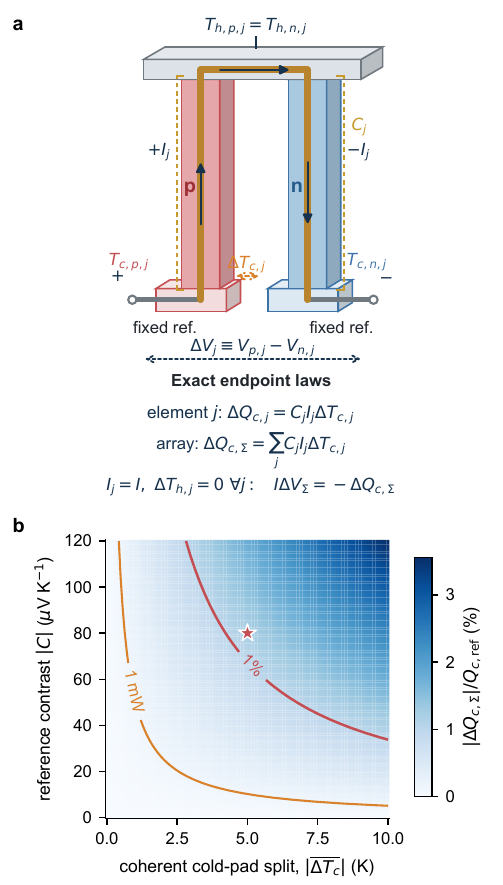}
 \caption{\label{fig:topology}\textbf{Endpoint topology exposes an exact
 \mbox{heat--voltage} relation.} (a) A split-pad element with an isothermal hot
 pair and electrically isolated, separately thermalized cold stages.  The
 leg-to-reference contrast $C_j$, current $I_j$, and signed split
 $\Delta T_{c,j}=T_{c,p,j}-T_{c,n,j}$ define the per-element endpoint response.  The dotted
 arrow denotes a high-impedance per-element measurement with
 $\Delta V_j=V_{p,j}-V_{n,j}$; series increments sum to $\Delta V_\Sigma$.
 (b) Magnitude of the aggregate cold-side increment relative to the modeled
 reference cooling heat.  The orange and red contours denote 1 mW and 1\%,
 respectively; they are model scales rather than detection or significance
 thresholds.  The star marks $C=+80~\mu$V K$^{-1}$, $I=+2.81$ A,
 $\overline{\Delta T_c}=+5$ K, and $\overline{\Delta T_h}=0$, giving 7.87 mW
 and $-2.80$ mV.}
\end{figure}

Using the published seven-pair operating point in
Ref.~\onlinecite{liu2026pbse} only as a normalization, $C=+80\,\mu$V K$^{-1}$,
$I=+2.81$ A, a coherent mean cold-pad split
$\overline{\Delta T_c}=+5$ K, and $\overline{\Delta T_h}=0$ give
$\Delta Q_{c,\Sigma}=7.87$ mW, or 1.19\% of the reference cooling heat,
together with $\Delta V_\Sigma=-2.80$ mV (Fig.~\ref{fig:topology}).  The 1-mW
and 1\% contours are deterministic model scales, not detection limits or
significance thresholds.  Current reversal
separates response parity but does not by itself remove current-odd Peltier,
Thomson, contact, and thermal-feedback backgrounds.  A direct test therefore
requires calibrated reference contrasts, endpoint temperatures, and simultaneous
heat and high-impedance voltage measurements.  If $u_Q$ and $u_V$ denote
independent standard uncertainties of the differential heat and voltage channels,
a 10\%-precision test at this representative point requires
$[u_Q^2+(Iu_V)^2]^{1/2}<0.79$ mW.  Equal uncertainty allocation gives
$u_Q<0.56$ mW and $u_V<0.20$ mV, making differential calorimetry and endpoint
stability the limiting requirements.

\begin{figure}[!t]
 \centering
 \includegraphics[width=\columnwidth]{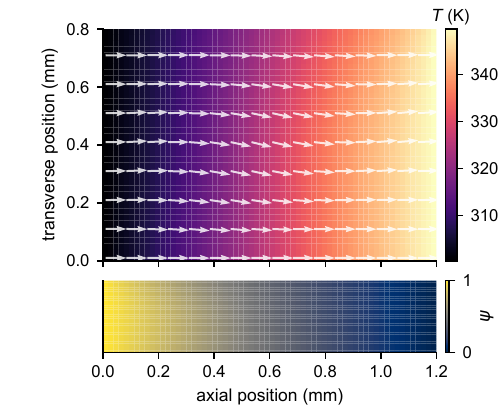}
 \caption{\label{fig:multidimensional}\textbf{Branch collection in a genuinely
 two-dimensional field.}  A heterogeneous temperature field and normalized
 current direction are shown above the corresponding cold-port adjoint collection
 field.  The transverse temperature span is 3.97 K and the maximum transverse-to-
 axial current-density ratio is 0.187.  The electrical resistivity and thermal
 conductivity vary spatially but are temperature independent in this calculation;
 the baseline Seebeck coefficient retains a temperature-dependent Thomson term.}
\end{figure}

We next assessed the multidimensional formulation with a conservative finite-
volume model containing heterogeneous electrical conductivity, transverse heat
flow, Joule heating, baseline Thomson transport, and a common-mode Thomson
perturbation (Fig.~\ref{fig:multidimensional}).  The representative field is
substantially two dimensional.  Twenty-seven parameter combinations and a
high-current case with an internal temperature maximum confirm the discrete
adjoint and the oriented-measure formulation; finite-difference agreement,
grid-convergence, and conservation results are given in the
supplementary material.  This calculation varies $\rho$ and $\kappa$ in space but
not with temperature.  It therefore tests multidimensional transfer and gradient
reversal.  The branch-collection response also persists in a separate synthetic,
fully coupled heterogeneous two-dimensional model with temperature-dependent
electrical and thermal transport, current-field redistribution, and the associated
Joule feedback.  Its medium-to-fine response change is 0.34\%; the governing
equations and convergence tests are given in the supplementary material.  All
reported reduced electrothermal Jacobians are nonsingular; this is the discrete
multidimensional counterpart of the one-dimensional baseline nondegeneracy
hypothesis.

Applications to published PbSe/Cr and Bi$_2$Te$_3$-based p/n transport curves
\cite{liu2026pbse,liu2025bts} give nominal net responses of a few milliwatts,
while the tested deterministic sensitivity envelopes extend into the 10-mW
range and can cross zero.  They also expose a data-identifiability limit: the
reported marginal accuracy bounds do not specify the joint p/n common-mode error
or its covariance.  These applications therefore establish conditional material
sensitivity rather than the sign of a specimen-specific response.  The processed
curves, perturbation sets, and separate first-order, fixed-current finite, and
reoptimized finite responses are reported in the supplementary material.

These results identify four device quantities:
$\alpha(T)$ sets differential thermoelectric drive, $\Gamma_{\rm cm}(T)$ describes
the common-mode Thomson content, $\mu_p-\mu_n$ determines its branch-to-port
collection, and endpoint
topology controls the direct term.  Electrical and thermal transport, geometry,
interfaces, and load remain required device parameters.  Equal Seebeck magnitude may
still be a practical proxy within related material families where these properties
covary, but it is not a network invariant or an independent design coordinate.  The
branch-measure equality condition extends existing theories of temperature-dependent
and graded-Seebeck cooling to paired thermoelectric branches.
\cite{snyder2012thomson,ryu2021three,zhao2026nonreciprocal}

The analysis assumes steady, scalar, local-equilibrium transport.  Finite common
thermal contacts enter through the network reduction, and the two-dimensional
calculations use fixed pad temperatures.  Tensorial transport, transients, active
shunts, and three-dimensional heat spreading within the package lie outside the
present steady-state scalar model.  A direct experimental test requires paired,
same-run $S_p(T)$ and $S_n(T)$ data with
covariance, calibrated endpoint contrasts, and simultaneous heat and voltage
measurements.  These observables determine whether a common-mode gradient is
transferred to the device port.

\section*{Supplementary Material}

See the supplementary material for complete derivations, finite-contact and
multidimensional tests, published-material reconstructions and sensitivity
analyses, endpoint-topology relations, and numerical convergence.

\begin{acknowledgments}
This work was supported by the National Science Fund for Distinguished Young
Scholars (No.~51925101), the National Natural Science Foundation of China
(Nos.~52450001 and 12104370), the Tianmushan Laboratory Research Project
(Nos.~TK2024D006 and TK2023C021), the ``Pioneer'' and ``Leading Goose'' R\&D
Program of Zhejiang (Project No.~2024SSYS0084), the Beijing Natural Science
Foundation (No.~JQ18004), the 111 Project (No.~B17002), and the Tencent Xplorer
Prize.  Additional support was provided by the Academic Excellence Foundation of
BUAA for PhD Students.  The authors acknowledge the high-performance computing
resources provided by Tianmushan Laboratory and Beihang University.
\end{acknowledgments}

\section*{Author Declarations}

\textbf{Conflict of Interest.} The authors have no conflicts to disclose.

\textbf{Author Contributions.} Peng Kang: Conceptualization, Methodology,
Software, Formal analysis, Investigation, Validation, Data curation,
Visualization, Writing--original draft, and Writing--review and editing. Da Wan,
Shulin Bai, Wei Yin, Peng Wang, Chenglong Wen, Zhen Li, and Yu Liu:
Investigation, Validation, and Writing--review and editing. Lei Zheng:
Conceptualization, Methodology, Supervision, Project administration, Funding
acquisition, and Writing--review and editing. Li-Dong Zhao: Conceptualization,
Supervision, Resources, Project administration, Funding acquisition, and
Writing--review and editing. All authors discussed the results and contributed to
the scientific interpretation and revision of the manuscript.

\section*{Data Availability}

The final versioned package of processed data, numerical outputs, and scripts
supporting this work will be released with the version of record when the
article is published in a peer-reviewed journal. Materials required for
editorial and referee assessment are available from the corresponding authors
during review. Publisher-hosted source articles used for figure digitization
remain available through the cited references and are not redistributed.

\bibliography{references}

\end{document}


\begin{center}
 {\large\bfseries Supplementary Material for ``Exact branch-transfer criterion
 for common-mode Thomson heat cancellation in thermoelectric couples''\par}
 \vspace{0.7em}
 {Peng Kang\textsuperscript{1,2,3}, Da Wan\textsuperscript{1,2,3},
 Shulin Bai\textsuperscript{1,2,3,4}, Wei Yin\textsuperscript{3},
 Peng Wang\textsuperscript{1,2}, Chenglong Wen\textsuperscript{1,2},
 Zhen Li\textsuperscript{1,2,3}, Yu Liu\textsuperscript{1,2,3},
 Lei Zheng\textsuperscript{1,2,3,a}, and
 Li-Dong Zhao\textsuperscript{1,2,3,4,b}\par}
 \vspace{0.45em}
 {\itshape\textsuperscript{1}School of Materials Science and Engineering,
 Beihang University, No.~37 Xueyuan Road, Beijing 100191, China\par}
 {\itshape\textsuperscript{2}National Key Laboratory of Artificial Intelligence
 for Material Science, Beihang University, No.~37 Xueyuan Road,
 Beijing 100191, China\par}
 {\itshape\textsuperscript{3}Tianmushan Laboratory, Beihang University,
 Hangzhou 311115, China\par}
 {\itshape\textsuperscript{4}Center for Bioinspired Science and Technology,
 Hangzhou International Innovation Institute, Beihang University,
 Hangzhou 311115, China\par}
 {\textsuperscript{a}zhenglei@buaa.edu.cn;
  \textsuperscript{b}zhaolidong@buaa.edu.cn\par}
\end{center}
\vspace{0.8em}

This Supplementary Material establishes the local and network transformations,
derives the fixed-current branch-to-port theorem, and presents the numerical
calculations and published-data case studies used in the Letter.  The central
distinction is between a global network co-shift, which transforms every
electrically active segment, and a p/n branch-relative perturbation against fixed
leads or interconnects.
Only the latter produces the endpoint and collection-measure response derived
below.

\section{Notation, sign convention, and the two common-mode operations}

Each branch is parameterized from its cold-side endpoint, $x=0$, to its
hot-side endpoint, $x=L_i$.  The local current $I_i$ is signed in the positive
$x$ direction, so $I_p=+I$ and $I_n=-I$ for a series p/n couple.  The branch
area is $A_i$, $K_i(T)=\kappa_i(T)A_i$, and
$r_i(T)=\rho_i(T)/A_i$.  The electric potential per unit charge is $\phi_i$,
the total heat rate in the positive-$x$ direction is $H_i$, and primes denote
$x$ derivatives unless an argument explicitly indicates differentiation with
respect to temperature.

We use the thermopower coordinates
\begin{equation}
 \alpha(T)=S_p(T)-S_n(T),\qquad
 M(T)=\frac{S_p(T)+S_n(T)}{2},\qquad
 \Gamma_{\mathrm{cm}}(T)=T\frac{\mathrm dM}{\mathrm dT}.
 \label{eq:s_coordinates}
\end{equation}
Here $\alpha$ is the differential thermopower that enters the open-circuit
p/n voltage, $M$ is the material common-mode coordinate, and
$\Gamma_{\mathrm{cm}}$ is its associated common-mode Thomson coefficient.
$\Gamma_{\mathrm{cm}}$ has units of thermopower and is not the dimensionless
generalized Thomson parameter used in some module theories.
The Kelvin relation connecting $T\,\mathrm dS/\mathrm dT$ to distributed
thermoelectric heat is established physics; the present result concerns how
that distributed heat is collected by a paired device port.
\cite{apertet2016kelvin,snyder2012thomson,ryu2021three}

Two operations that look similar on isolated p/n curves are physically
different.  A \emph{global network co-shift} adds a temperature-independent
constant $C$ to every electrically active modeled segment, including
reference leads and interconnects.  A \emph{branch-relative common
perturbation} adds the same function $\epsilon m(T)$ to the p and n
semiconductor branches while the reference and interconnect thermopowers are
held fixed.  The second operation preserves $\alpha(T)$ pointwise but changes
material-to-reference contrasts and, when $m$ varies with temperature, changes
the branch Thomson coefficients by $\epsilon\Gamma_m(T)$, where
$\Gamma_m=T\,\mathrm dm/\mathrm dT$.  This perturbation profile is distinct
from the baseline material coordinate
$\Gamma_{\mathrm{cm}}=T\,\mathrm dM/\mathrm dT$ in
Eq.~(\ref{eq:s_coordinates}).

\begin{table}[htbp]
\caption{Physical distinction between the two common-mode operations.}
\label{tab:s_operations}
\small
\begin{tabular}{llll}
\toprule
\sftcell{0.17\textwidth}{Operation} &
\sftcell{0.19\textwidth}{Transformed object} &
\sftcell{0.21\textwidth}{First-order source} &
\sftcell{0.26\textwidth}{Null condition} \\
\midrule
\sftcell{0.17\textwidth}{Global network co-shift $C$} &
\sftcell{0.19\textwidth}{Every electrically active segment and reference path} &
\sftcell{0.21\textwidth}{No distributed term; local heat changes by $CTI$} &
\sftcell{0.26\textwidth}{Nodewise current conservation and a closed electrical path} \\
\sftcell{0.17\textwidth}{Branch-relative $m(T)$} &
\sftcell{0.19\textwidth}{p/n branches only, against fixed leads or interconnects} &
\sftcell{0.21\textwidth}{Endpoint Peltier term plus $\Gamma_m=T\,\mathrm dm/\mathrm dT$} &
\sftcell{0.26\textwidth}{Vanishing endpoint term and equality, or profile-specific orthogonality, of collection measures} \\
\bottomrule
\end{tabular}
\end{table}

\section{Exact local constitutive transformation and network null}

For steady scalar local-equilibrium transport, the constitutive and energy
relations can be written
\begin{align}
 \phi_i'&=-\frac{\rho_i I_i}{A_i}-S_iT_i',
 \label{eq:s_phi}\\
 H_i&=S_iT_iI_i-K_iT_i',
 \label{eq:s_heat}\\
 H_i'&=I_i(-\phi_i').
 \label{eq:s_energy}
\end{align}
Eliminating $H_i$ and $\phi_i$ yields
\begin{equation}
 \frac{\mathrm d}{\mathrm dx}\!\left[K_i(T_i)T_i'\right]
 +r_i(T_i)I_i^2-\tau_i(T_i)I_iT_i'=0,
 \qquad \tau_i(T)=T\frac{\mathrm dS_i}{\mathrm dT}.
 \label{eq:s_temperature}
\end{equation}
These equations retain Peltier transport in $H_i$, Joule heating through
$r_iI_i^2$, and Thomson redistribution through $\tau_iI_iT_i'$ without
double-counting any term.

Under a global constant co-shift, define
\begin{equation}
 S_i^*=S_i+C,\qquad
 \phi_i^*=\phi_i-C(T_i-T_{\mathrm{ref}}),\qquad
 H_i^*=H_i+CT_iI_i.
 \label{eq:s_global_transform}
\end{equation}
Substitution into Eqs.~(\ref{eq:s_phi})--(\ref{eq:s_energy}) shows pointwise
invariance with the same temperature and current fields.  At any electrical
network node $a$ with temperature $T_a$, currents directed consistently away
from the node satisfy $\sum_{j\in a}I_j=0$, and therefore
\begin{equation}
 \sum_{j\in a}\Delta H_j=CT_a\sum_{j\in a}I_j=0.
 \label{eq:s_node_null}
\end{equation}
The voltage increment $-C\,\mathrm dT$ is an exact differential, so its
closed-loop integral also vanishes.  Thus the complete terminal
$I$--$V$--$Q$ map and the steady node solution are unchanged.  The result is
independent of whether different network nodes have different temperatures;
the cancellation is local to each node.  It does not assert that absolute
thermopower is unmeasurable, because absolute-Seebeck metrology compares a
sample to an explicitly realized reference and therefore does not perform the
global transformation of Eq.~(\ref{eq:s_global_transform}).
\cite{roberts1977absolute,amagai2019absolute}

The transformation must include the complete electrical network.  An electrically conducting shunt, thermoelectric
contact layer, or reference lead that is omitted from the transformation can
leave an interface contrast.  A passive heat leak carries no current and
requires no shift.  Finite thermal contacts between shared nodes and reservoirs
do not break the null because the bulk terminal heats and node balances remain
unchanged.  Magnetic thermoelectric tensors, transient storage, and nonlocal
transport require a correspondingly enlarged constitutive operator.

\section{Linearized branch response and cold-port adjoint}

We now apply the branch-relative perturbation
\begin{equation}
 S_i^\epsilon(T)=S_i(T)+\epsilon m(T),\qquad
 \tau_i^\epsilon(T)=\tau_i(T)+\epsilon\Gamma_m(T),\qquad
 \Gamma_m(T)=T\frac{\mathrm dm}{\mathrm dT},
 \label{eq:s_branch_perturbation}
\end{equation}
at fixed $I_i$, fixed branch endpoint temperatures, fixed $\rho_i(T)$ and
$\kappa_i(T)$, and fixed geometry.  Write
$T_i^\epsilon=T_i+\epsilon y_i+O(\epsilon^2)$.  Differentiating
Eq.~(\ref{eq:s_temperature}) gives
\begin{equation}
 \mathcal L_i y_i=I_i\Gamma_m(T_i)T_i',\qquad
 y_i(0)=y_i(L_i)=0,
 \label{eq:s_linearized}
\end{equation}
with
\begin{align}
 \mathcal L_i y={}&
 \frac{\mathrm d}{\mathrm dx}
 \left[K_i y'+K_{i,T}T_i'y\right]
 +r_{i,T}I_i^2y \nonumber\\
 &-I_i\left(\tau_i y'+\tau_{i,T}T_i'y\right).
 \label{eq:s_linear_operator}
\end{align}
All coefficients in Eq.~(\ref{eq:s_linear_operator}) are evaluated on the
unperturbed branch.  The $r_{i,T}$, $K_{i,T}$, and $\tau_{i,T}$ terms are part
of the general first variation; omitting them would impose an additional
constant-property approximation.

Writing the operator as
$(a_iy')'+(b_iy)'+c_iy'+d_iy$, its formal adjoint is
$(a_i\psi')'-b_i\psi'-(c_i\psi)'+d_i\psi$.  Substitution produces exact
cancellation of the $K_{i,T}T_i'\psi_i'$ and
$I_i\tau_{i,T}T_i'\psi_i$ pairs, leaving
\begin{equation}
 \boxed{\mathcal L_i^\dagger\psi_i
 =K_i(T_i)\psi_i''+I_i\tau_i(T_i)\psi_i'
 +r_{i,T}(T_i)I_i^2\psi_i=0.}
 \label{eq:s_adjoint}
\end{equation}
For cold-port collection, the boundary conditions are
\begin{equation}
 \psi_i(0)=1,\qquad \psi_i(L_i)=0.
 \label{eq:s_adjoint_bc}
\end{equation}

For positive $K_i$ and bounded continuous coefficients, the response theorem
is formulated about a nondegenerate baseline: the homogeneous Dirichlet
problem associated with $\mathcal L_i$ has only the trivial solution.
Equivalently for this regular second-order boundary-value problem, the
homogeneous adjoint Dirichlet problem has no zero mode.  Define
\begin{equation}
 \mathfrak a_i(x)=\frac{I_i\tau_i[T_i(x)]}{K_i[T_i(x)]},\qquad
 \mathfrak q_i(x)=\frac{I_i^2r_{i,T}[T_i(x)]}{K_i[T_i(x)]},\qquad
 P_i(x)=\exp\!\left[\int_0^x \mathfrak a_i(s)\,\mathrm ds\right].
 \label{eq:s_nondegenerate_weights}
\end{equation}
The homogeneous adjoint equation is then
$(P_i\psi_i')'+P_i\mathfrak q_i\psi_i=0$.  Multiplication by $\psi_i$,
integration over $[0,L_i]$, and the Dirichlet Poincar\'e inequality show that a
zero mode is excluded when
\begin{equation}
 \boxed{\eta_i\equiv
 \frac{L_i^2}{\pi^2}\frac{P_{i,\max}}{P_{i,\min}}
 \left\|(\mathfrak q_i)_+\right\|_\infty<1,}
 \qquad (\mathfrak q_i)_+=\max(\mathfrak q_i,0).
 \label{eq:s_nondegenerate_condition}
\end{equation}
Indeed, any homogeneous solution would satisfy
\begin{equation}
 \int_0^{L_i}P_i(\psi_i')^2\,\mathrm dx
 =\int_0^{L_i}P_i\mathfrak q_i\psi_i^2\,\mathrm dx
 \leq \eta_i\int_0^{L_i}P_i(\psi_i')^2\,\mathrm dx,
 \label{eq:s_nondegenerate_energy}
\end{equation}
which forces $\psi_i=0$ when $\eta_i<1$.  If $\mathfrak q_i\leq0$,
uniqueness follows without the bound.  Equation~(\ref{eq:s_nondegenerate_condition})
is sufficient rather than necessary;
outside it, invertibility can instead be checked directly for the linearized
boundary-value operator.

We evaluated Eq.~(\ref{eq:s_nondegenerate_condition}) for 2102 leg baselines
from 1051 PbSe/Cr and Bi$_2$Te$_3$-based couple states, including the current scans,
Seebeck perturbation sets, transport-property corners, finite-difference
states, and thermal-contact endpoint states used below.  Every state satisfies
the sufficient condition.  Representative values are given in Table
\ref{tab:s_nondegenerate}; the global maximum, $\eta_i=0.998402$, occurs only at
the extreme $-3.5$ A PbSe/Cr p-leg state.  As an independent continuous-
coefficient check in the normalized coordinate $u=x/L_i$, the minimum absolute
shooting determinant over the same set is 0.361376.  Proximity of $\eta_i$ to
unity marks near-exhaustion of this sufficient coercivity certificate, not
approach to a zero mode: the $-3.5$ A state has an absolute shooting determinant
of 0.611820, while the minimum value 0.361376 occurs at a different $+3.5$ A
state.  These results establish
nondegeneracy for the states used here, not for arbitrary currents or material
laws.

\begin{table}[htbp]
\caption{Weighted Poincar\'e ratios for the reported material-model states.
Values below unity satisfy Eq.~(\ref{eq:s_nondegenerate_condition}).}
\label{tab:s_nondegenerate}
\small
\begin{tabular}{lcc}
\toprule
State & $\eta_p$ & $\eta_n$ \\
\midrule
PbSe/Cr, nominal original, $I=2.81$ A & 0.387820 & 0.196954 \\
PbSe/Cr, common-mode flattened, $I=2.81$ A & 0.364280 & 0.210627 \\
Bi$_2$Te$_3$-based reference, $I=3$ A & 0.031976 & 0.040414 \\
All evaluated states & \multicolumn{2}{c}{maximum $\eta_i=0.998402$} \\
\bottomrule
\end{tabular}
\end{table}

Green's identity, Eq.~(\ref{eq:s_linearized}), and the homogeneous endpoint
conditions on $y_i$ then give
\begin{equation}
 -K_i(T_{c,i})y_i'(0)
 =\int_0^{L_i}\psi_i(x)I_i\Gamma_m[T_i(x)]T_i'(x)\,\mathrm dx.
 \label{eq:s_green}
\end{equation}
Since $H_{c,i}=S_i(T_{c,i})T_{c,i}I_i-K_i(T_{c,i})T_i'(0)$,
the complete first-order cold-end response of one branch is
\begin{equation}
 \boxed{\delta H_{c,i}=I_iT_{c,i}m(T_{c,i})
 +\int_0^{L_i}\psi_i I_i\Gamma_m(T_i)T_i'\,\mathrm dx.}
 \label{eq:s_branch_response}
\end{equation}
The first term is a direct boundary Peltier contrast and the second is the
distributed source filtered by the cold-port collection field.  General
thermoelectric adjoint methods are established; Eqs.~(\ref{eq:s_adjoint}) and
(\ref{eq:s_branch_response}) specialize that logic to a common p/n Seebeck
perturbation and a specified heat port.
\cite{lundgaard2018adjoint,reales2024adjoint}

\section{Oriented temperature-space measure theorem}

For any Borel set $B$ in a compact positive-temperature interval containing
the branch fields, define the oriented push-forward measure
\begin{equation}
 \mu_i(B)=\int_0^{L_i}
 \mathbf 1_B[T_i(x)]\,\psi_i(x)T_i'(x)\,\mathrm dx.
 \label{eq:s_measure}
\end{equation}
This is a finite signed measure, not a probability distribution.  For a
strictly increasing temperature field it has density
$\psi_i[x_i(T)]$ with respect to $\mathrm dT$.  For a nonmonotonic field and
regular values of $T$, the density is instead
\begin{equation}
 \frac{\mathrm d\mu_i}{\mathrm dT}
 =\sum_{x:\,T_i(x)=T}\psi_i(x)\,\operatorname{sgn}T_i'(x),
 \label{eq:s_oriented_density}
\end{equation}
so repeated visits to the same temperature retain their orientation.  The
measure definition in Eq.~(\ref{eq:s_measure}) remains valid at internal
turning points where the inverse $x_i(T)$ does not.

Summing Eq.~(\ref{eq:s_branch_response}) for $I_p=+I$ and $I_n=-I$ gives the
pair theorem
\begin{equation}
 \boxed{
 \delta Q_c=
 I\left[T_{c,p}m(T_{c,p})-T_{c,n}m(T_{c,n})\right]
 +I\int\Gamma_m(T)\,\mathrm d(\mu_p-\mu_n).}
 \label{eq:s_pair_theorem}
\end{equation}
The first bracket is the endpoint term.  It vanishes at a shared isothermal
cold node, and both cold and hot endpoint terms vanish in the corresponding
full-port statements for a shared-isothermal p/n topology.  In the monotonic
shared-endpoint special case, Eq.~(\ref{eq:s_pair_theorem}) becomes
\begin{equation}
 \delta Q_c=I\int_{T_c}^{T_h}\Gamma_m(T)
 \left\{\psi_p[x_p(T)]-\psi_n[x_n(T)]\right\}\,\mathrm dT.
 \label{eq:s_monotone_kernel}
\end{equation}

Equation~(\ref{eq:s_pair_theorem}) yields two distinct cancellation tests.
For one specified profile $\Gamma_m$, only orthogonality is required:
$\int\Gamma_m\,\mathrm d(\mu_p-\mu_n)=0$.  If the endpoint term is zero, the
response vanishes for every continuous admissible $\Gamma_m$ if and only if
\begin{equation}
 \boxed{\mu_p=\mu_n.}
 \label{eq:s_measure_equality}
\end{equation}
Sufficiency follows immediately from Eq.~(\ref{eq:s_pair_theorem}).  For the
converse, vanishing against every continuous $\Gamma_m$ implies equality of the
finite signed measures by uniqueness in the Riesz representation theorem.
Every continuous $\Gamma_m$ on a positive-temperature interval is represented by
the anchored perturbation
$m(T)=\int_{T_0}^{T}\Gamma_m(\vartheta)/\vartheta\,\mathrm d\vartheta$, so the
construction spans the full continuous function class.  Matching unsigned branch
material functions is not by itself sufficient, because the
thermoelectric terms also contain the signed branch currents.  A sufficient
condition is equality of the oriented baseline and adjoint operators (for example,
matched $K$ and $r$ with $I_p\tau_p=I_n\tau_n$ on the same coordinate domain and
under the same endpoint boundary conditions); this condition is not necessary,
because nonidentical branches can still be transfer matched.

The theorem is a fixed-current, first-order statement.  A finite-amplitude
perturbation changes the temperature fields and collection measures,
and an optimized-current comparison also moves the operating point.  These
observables are reported separately below rather than being interpreted as the
same derivative.

\section{Exact constant-property reduction: leading $I^3$ and finite-current $I^5$ terms}

Consider constant $\rho_i$ and $\kappa_i$, a constant-Seebeck baseline,
uniform area, and fixed common endpoints.  On $u=x/L_i$, define the
Joule-curvature temperature
\begin{equation}
 g_i=\frac{\rho_iI^2L_i^2}{2\kappa_iA_i^2}
 =\frac{I^2}{2}\frac{R_i}{K_i^{\mathrm{leg}}},
 \quad R_i=\frac{\rho_iL_i}{A_i},\quad
 K_i^{\mathrm{leg}}=\frac{\kappa_iA_i}{L_i}.
 \label{eq:s_g_definition}
\end{equation}
The baseline temperature and cold-port adjoint fields are exactly
\begin{equation}
 T_i(u)=T_c+\Delta T\,u+g_i u(1-u),\qquad
 \psi_i(u)=1-u.
 \label{eq:s_constant_fields}
\end{equation}
For the anchored linear common mode $m(T)=b(T-T_c)$,
$\Gamma_m(T)=bT$ and the direct endpoint term is zero.  For either branch,
integration by parts gives
\begin{align}
 Ib\int_0^1(1-u)T_i\frac{\mathrm dT_i}{\mathrm du}\,\mathrm du
 &=\frac{Ib}{2}\int_0^1(1-u)\,\mathrm d(T_i^2)\nonumber\\
 &=\frac{Ib}{2}\left([(1-u)T_i^2]_0^1
 +\int_0^1T_i^2\,\mathrm du\right).
 \label{eq:s_constant_parts}
\end{align}
The boundary contribution is $-IbT_c^2/2$ for each branch and cancels in the
p--n difference at the shared cold endpoint.  The pair response is therefore
\begin{equation}
 \delta Q_c=\frac{Ib}{2}\int_0^1
 \left[T_p(u)^2-T_n(u)^2\right]\,\mathrm du.
 \label{eq:s_constant_integral}
\end{equation}
Using
$\int_0^1u(1-u)\,\mathrm du=1/6$,
$\int_0^1u^2(1-u)\,\mathrm du=1/12$, and
$\int_0^1u^2(1-u)^2\,\mathrm du=1/30$ gives
\begin{equation}
 \boxed{\delta Q_c=Ib(g_p-g_n)
 \left[\frac{T_c}{6}+\frac{\Delta T}{12}
 +\frac{g_p+g_n}{60}\right].}
 \label{eq:s_constant_closed_form}
\end{equation}
Because $g_i\propto I^2$, the first two bracketed terms yield a leading
$I^3b[R_p/K_p^{\mathrm{leg}}-R_n/K_n^{\mathrm{leg}}]$ response, while the last
term is an exact $I^5$
correction.  The law is therefore cubic only in the low-Joule limit.  It also
shows why raw branch identity is unnecessarily restrictive: any nonidentical
pair with matched $R_i/K_i^{\mathrm{leg}}$ has $g_p=g_n$ and cancels for this
entire linear-$m$ family.

Normalizing by the cold Peltier scale $\alpha I T_c$ gives
\begin{equation}
 \frac{\delta Q_c}{\alpha I T_c}
 =\mathcal M\,\Delta G
 \left(\frac{1}{6}+\frac{\theta}{12}+\frac{\bar G}{30}\right),
 \label{eq:s_dimensionless}
\end{equation}
where
$\mathcal M=bT_c/\alpha$,
$\Delta G=(g_p-g_n)/T_c$,
$\bar G=(g_p+g_n)/(2T_c)$, and
$\theta=\Delta T/T_c$.  Equation~(\ref{eq:s_dimensionless}) is a local
near-symmetry coordinate, not a replacement for the full variable-property
measure in Eq.~(\ref{eq:s_pair_theorem}).

\subsection{Closed-form, adjoint, and nonlinear finite-difference comparison}

The reference constant-property case used
$T_c=\SI{300}{\kelvin}$, $T_h=\SI{330}{\kelvin}$,
$I=\SI{0.5}{\ampere}$, $b=\SI{1e-6}{\volt\per\kelvin\squared}$,
$L_p=L_n=\SI{1}{\milli\meter}$, and
$A_p=A_n=\SI{1}{\milli\meter\squared}$.  The p leg used
$(\rho_p,\kappa_p)=(\SI{1e-5}{\ohm\meter},
\SI{1.5}{\watt\per\meter\per\kelvin})$ and the n leg used
$(\rho_n,\kappa_n)=(\SI{2e-5}{\ohm\meter},
\SI{1.0}{\watt\per\meter\per\kelvin})$.
Equation~(\ref{eq:s_constant_closed_form}) gives
$\delta Q_c=-\SI{43.80}{\micro\watt}$ per unit $\epsilon$.
The independently evaluated adjoint differs by
$2.40\times10^{-14}$ in relative terms.  Central differences of the full
nonlinear boundary-value problem converge to the same value (Table
\ref{tab:s_1d_verification}).

\begin{table}[htbp]
\caption{One-dimensional comparison for the constant-property response.}
\label{tab:s_1d_verification}
\small
\begin{tabular}{ccc}
\toprule
Central step $h$ & Nonlinear derivative ($\mu$W) & Relative error \\
\midrule
0.100 & $-43.7965828$ & $6.54\times10^{-6}$ \\
0.030 & $-43.7963221$ & $5.89\times10^{-7}$ \\
0.010 & $-43.7962992$ & $6.57\times10^{-8}$ \\
0.003 & $-43.7962966$ & $6.11\times10^{-9}$ \\
0.001 & $-43.7962963$ & $9.48\times10^{-10}$ \\
\bottomrule
\end{tabular}
\end{table}

An identical-leg control gives zero exactly.  A second control changes the
individual geometry and transport coefficients while preserving
$R_p/K_p^{\mathrm{leg}}=R_n/K_n^{\mathrm{leg}}$; its maximum collection-field
difference is $2.22\times10^{-16}$ and its closed-form response is
$5.83\times10^{-21}$ W.  This control distinguishes transfer matching from
mere equality of raw leg parameters.

\section{Finite thermal contacts as an endpoint Schur complement}

The fixed-endpoint derivative can be coupled to an external thermal network
without treating the contacts as distributed two-dimensional regions.  Let
$\bm z=(T_c^{\mathrm{leg}},T_h^{\mathrm{leg}})^T$ be the common p/n bulk
endpoint temperatures and
$\bm q(\bm z,\epsilon)=(Q_c,Q_h)^T$ the bulk port heats.  For reservoir
temperatures $T_{c,r}$ and $T_{h,r}$ and contact resistances
$R_{\mathrm{th},c}$ and $R_{\mathrm{th},h}$, write the two node residuals
\begin{equation}
 \bm F(\bm z,\epsilon)=
 \begin{pmatrix}
 T_c^{\mathrm{leg}}-T_{c,r}+R_{\mathrm{th},c}Q_c\\
 T_h^{\mathrm{leg}}-T_{h,r}-R_{\mathrm{th},h}Q_h
 \end{pmatrix}=\bm 0.
 \label{eq:s_contact_residual}
\end{equation}
Differentiation gives
\begin{equation}
 \bm z_\epsilon=-\bm F_{\bm z}^{-1}\bm F_\epsilon,
 \qquad
 \frac{\mathrm dQ_c}{\mathrm d\epsilon}
 =Q_{c,\epsilon}+\nabla_{\bm z}Q_c\cdot\bm z_\epsilon.
 \label{eq:s_schur}
\end{equation}
Thus the fixed-endpoint kernel supplies $\bm q_\epsilon$, and the contact
network supplies the endpoint-temperature feedback.  At fixed current,
series-contact Joule heat, its prescribed cold/hot partition, and a passive
reservoir-to-reservoir conductance have no direct common-mode derivative, but
they can change the baseline operating point and any optimized-current result.

The finite-contact reference model used
$(R_{\mathrm{th},c},R_{\mathrm{th},h})=(0.8,0.5)$ K W$^{-1}$,
$R_c=\SI{0.01}{\ohm}$, a 0.4 cold-side Joule fraction, a parasitic
conductance of \SI{2e-4}{\watt\per\kelvin}, and $I=\SI{0.5}{\ampere}$.
The contact-coupled derivative was
$-\SI{43.65}{\micro\watt}$, compared with
$-\SI{43.65}{\micro\watt}$ from a central difference of the full
boundary network, a relative difference of $6.55\times10^{-8}$.

\begin{table}[htbp]
\caption{Finite-contact endpoint reduction for the constant-property reference case.}
\label{tab:s_contact}
\small
\begin{tabular}{lr}
\toprule
\sftcell{0.49\textwidth}{Quantity} & Value \\
\midrule
\sftcell{0.49\textwidth}{Baseline $(T_c^{\mathrm{leg}},T_h^{\mathrm{leg}})$} & $(300.015762,329.998147)$ K \\
\sftcell{0.49\textwidth}{Fixed-endpoint $(Q_{c,\epsilon},Q_{h,\epsilon})$} & $(-43.7972622,-43.7972622)$ $\mu$W \\
\sftcell{0.49\textwidth}{Induced $(T_{c,\epsilon},T_{h,\epsilon})$} & $(+34.9187,-21.8299)$ $\mu$K \\
\sftcell{0.49\textwidth}{Schur-complement $\mathrm dQ_c/\mathrm d\epsilon$} & $-43.6484069$ $\mu$W \\
\sftcell{0.49\textwidth}{Full-network central difference} & $-43.6484098$ $\mu$W \\
\sftcell{0.49\textwidth}{Relative difference} & $6.55\times10^{-8}$ \\
\bottomrule
\end{tabular}
\end{table}

This calculation confirms the network reduction in Eq.~(\ref{eq:s_schur}).
The two-dimensional formulation below retains fixed-temperature end pads and
varies sidewall coupling.

\section{Two-dimensional conservation model}

Here $k\equiv\kappa$ denotes thermal conductivity.

The two-dimensional equations were solved with a finite-volume formulation
developed separately from the one-dimensional model.  Each p or n branch is a rectangular finite-volume
domain with spatially heterogeneous, temperature-independent $\rho(x,y)$ and
$k(x,y)$; $S(T)$ and its Thomson coefficient retain their temperature
dependence.  Thus this model tests transverse current flow, spatial property
contrast, sidewall loss, and nonmonotonic temperature fields, but not the
additional electrothermal feedback from $\rho_T$ or $k_T$.  The electrical
problem uses the electrochemical potential
\begin{equation}
 \nabla\!\cdot[\sigma(x,y)\nabla\Psi]=0,\qquad
 \bm J=-\sigma\nabla\Psi,\qquad
 \Psi=\phi+\int^T S(\vartheta)\,\mathrm d\vartheta,
 \label{eq:s_2d_electrical}
\end{equation}
with insulated sidewalls.  A unit-potential solution is linearly scaled to the
specified signed branch current.  Joule heating on each electrical face is
partitioned conservatively between adjacent control volumes, or assigned to the
single adjacent volume at a boundary.  The thermal problem is
\begin{equation}
 \nabla\!\cdot(k\nabla T)+\rho|\bm J|^2
 -\tau(T)\bm J\!\cdot\nabla T=0,
 \label{eq:s_2d_thermal}
\end{equation}
with Dirichlet cold/hot end temperatures and asymmetric Robin side loss.

For any common-mode perturbation basis, let $G'(T)=\Gamma_m(T)$, and let
$G_0'(T)=\tau_0(T)$ denote the baseline branch Thomson coefficient.  Current
conservation gives
\begin{equation}
 \Gamma_m(T)\bm J\!\cdot\nabla T
 =\nabla\!\cdot[\bm JG(T)],
 \label{eq:s_2d_flux}
\end{equation}
so the perturbation Thomson term was assembled as a conservative flux rather
than a cellwise nonconservative source.  In discrete form the residual is
\begin{equation}
 \bm R(\bm T,\epsilon)=\bm A\bm T-\bm b
 +\bm D\bm G_0(\bm T)+\epsilon\bm D\bm G(\bm T)=\bm0.
 \label{eq:s_2d_residual}
\end{equation}
The full base tangent is
$\bm L=\bm A+\bm D\operatorname{diag}[\tau_0(\bm T)]$.
If $\bm c$ collects the cold conductive-flux weights, the discrete collection
field solves
\begin{equation}
 \bm L^T\bm\psi=\bm c,
 \qquad
 \delta Q_c=\delta Q_{c,\mathrm{end}}
 +\bm\psi^T\bm D\bm G(\bm T).
 \label{eq:s_2d_adjoint}
\end{equation}
This is the finite-volume counterpart of
Eq.~(\ref{eq:s_branch_response}).  It includes the transpose of the full
base-Thomson tangent, not only the symmetric conduction matrix.

\subsection{Heterogeneous transport fields and parameter space}

With normalized
coordinates $\xi=x/L$ and $\eta=y/W$, the fields were
\begin{align}
 \rho/\rho_0={}&\exp\{0.33\sin(2\pi\xi+\varphi)\cos(\pi\eta)
 +0.17(\eta-0.5)\nonumber\\
 &\hspace{2.6em}+0.11\sin(3\pi\xi-2\pi\eta+0.4+\varphi)\},
 \label{eq:s_rho_pattern}\\
 k/k_0={}&1+0.27\cos(2\pi\xi+0.3+\varphi)\sin(\pi\eta)
 +0.16(\eta-0.5)\nonumber\\
 &\hspace{2.6em}+0.09\sin(\pi\xi+2\pi\eta-\varphi).
 \label{eq:s_k_pattern}
\end{align}
The three mismatch families are defined in Table~\ref{tab:s_2d_design}.
All used $S_p(300\,\mathrm K)=+215~\mu$V K$^{-1}$ and
$S_n(300\,\mathrm K)=-185~\mu$V K$^{-1}$.  The heterogeneity phases
$(\varphi_p,\varphi_n)$ were $(0.20,0.62)$, $(0.15,1.05)$, and
$(0.35,1.35)$ for the near-matched, property-contrast, and
geometry--property-contrast families, respectively.

\begin{table*}[htbp]
\caption{Base parameters of the three two-dimensional mismatch families at
unit sidewall multiplier.}
\label{tab:s_2d_design}
\small
\begin{tabular}{lllllll}
\toprule
\sftcell{0.14\textwidth}{Family} & Leg &
\sftcell{0.13\textwidth}{$(L,W,D)$ (mm)} &
\sftcell{0.12\textwidth}{$\rho_0$ ($10^{-5}\,\Omega\,\mathrm m$)} &
\sftcell{0.09\textwidth}{$k_0$ (W m$^{-1}$ K$^{-1}$)} &
\sftcell{0.10\textwidth}{$\mathrm dS/\mathrm dT$ ($\mu$V K$^{-2}$)} &
\sftcell{0.23\textwidth}{$(h_b,h_t)$ (W m$^{-2}$ K$^{-1}$); $(T_{\infty,b},T_{\infty,t})$ (K)} \\
\midrule
\sftcell{0.14\textwidth}{Near matched} & p & $(1.20,0.80,0.80)$ & 1.30 & 1.18 & $+0.45$ & \sftcell{0.23\textwidth}{$(100,480)$; $(316,294)$} \\
 & n & $(1.20,0.80,0.80)$ & 1.45 & 1.08 & $-0.32$ & \sftcell{0.23\textwidth}{$(115,450)$; $(315,295)$} \\
\sftcell{0.14\textwidth}{Property contrast} & p & $(1.20,0.80,0.80)$ & 1.18 & 1.42 & $+0.55$ & \sftcell{0.23\textwidth}{$(95,570)$; $(317,293)$} \\
 & n & $(1.20,0.80,0.80)$ & 1.88 & 0.92 & $-0.38$ & \sftcell{0.23\textwidth}{$(145,410)$; $(314,296)$} \\
\sftcell{0.14\textwidth}{Geometry--property contrast} & p & $(1.48,0.62,0.72)$ & 1.02 & 1.62 & $+0.65$ & \sftcell{0.23\textwidth}{$(75,650)$; $(318,292)$} \\
 & n & $(0.84,0.98,0.66)$ & 2.25 & 0.71 & $-0.46$ & \sftcell{0.23\textwidth}{$(190,340)$; $(312,298)$} \\
\bottomrule
\end{tabular}
\end{table*}

The 27 cases combine the three mismatch families with sidewall
multipliers $0.25$, 1, and 4 and the following three explicitly defined
common-mode bases:
\begin{align}
 \Gamma_{m,1}(T)&=\gamma_0,\quad
 m_1(T)=\gamma_0\ln(T/T_0),\quad
 \gamma_0=\SI{3e-4}{\volt\per\kelvin},\nonumber\\
 \Gamma_{m,2}(T)&=bT,\quad
 m_2(T)=b(T-T_0),\quad
 b=\SI{1e-6}{\volt\per\kelvin\squared},\nonumber\\
 \Gamma_{m,3}(T)&=b_gT\exp[-(T-T_*)^2/(2w^2)],\quad
 b_g=\SI{1.4e-6}{\volt\per\kelvin\squared},\nonumber\\
 &\hspace{8em}T_*=\SI{330}{\kelvin},\quad w=\SI{12}{\kelvin}.
 \label{eq:s_2d_bases}
\end{align}
The modes were anchored at $T_0=\SI{300}{\kelvin}$.  The basis amplitudes are
different and sample distinct flux shapes.  Comparisons across the three bases
therefore concern adjoint--finite-difference agreement rather than response
magnitude.  Every case used
$T_c=\SI{300}{\kelvin}$, $T_h=\SI{350}{\kelvin}$,
$|I|=\SI{1.1}{\ampere}$, a $24\times16$ grid per branch, and an independent
central-difference step $\epsilon=10^{-3}$.

\subsection{Two-dimensional response and grid convergence}

Table~\ref{tab:s_2d_matrix} reports the range across the three sidewall
multipliers for each mismatch--basis pair.  The worst adjoint--finite-
difference relative difference was $3.96\times10^{-8}$, and the worst global
energy residual was $5.45\times10^{-15}$ W.  The branch energy convention was
$Q_h-Q_c+Q_{\mathrm{side}}=P_{\mathrm{electrical}}$.

\begin{table*}[htbp]
\caption{Summary of the 27 two-dimensional parameter combinations.}
\label{tab:s_2d_matrix}
\small
\begin{tabular}{lccc}
\toprule
\sftcell{0.25\textwidth}{Mismatch family and $\Gamma_m$ basis} &
\sftcell{0.21\textwidth}{Adjoint range (mW per $\epsilon$)} &
\sftcell{0.18\textwidth}{Maximum relative difference} &
\sftcell{0.19\textwidth}{Maximum energy residual (W)} \\
\midrule
\sftcell{0.25\textwidth}{Near matched / constant} & $[-0.42158,-0.21225]$ & $6.68\times10^{-9}$ & $3.29\times10^{-15}$ \\
\sftcell{0.25\textwidth}{Near matched / linear} & $[-0.46075,-0.22484]$ & $6.09\times10^{-9}$ & $3.05\times10^{-15}$ \\
\sftcell{0.25\textwidth}{Near matched / localized} & $[-0.45730,-0.17145]$ & $1.58\times10^{-8}$ & $3.18\times10^{-15}$ \\
\sftcell{0.25\textwidth}{Property contrast / constant} & $[-1.27477,-0.53457]$ & $9.44\times10^{-9}$ & $3.63\times10^{-15}$ \\
\sftcell{0.25\textwidth}{Property contrast / linear} & $[-1.39760,-0.56368]$ & $1.40\times10^{-8}$ & $3.76\times10^{-15}$ \\
\sftcell{0.25\textwidth}{Property contrast / localized} & $[-1.37838,-0.40300]$ & $3.96\times10^{-8}$ & $3.67\times10^{-15}$ \\
\sftcell{0.25\textwidth}{Geometry--property contrast / constant} & $[-2.92192,-0.48997]$ & $1.49\times10^{-8}$ & $5.45\times10^{-15}$ \\
\sftcell{0.25\textwidth}{Geometry--property contrast / linear} & $[-3.12550,-0.53463]$ & $1.92\times10^{-8}$ & $5.34\times10^{-15}$ \\
\sftcell{0.25\textwidth}{Geometry--property contrast / localized} & $[-2.65667,-0.52997]$ & $2.69\times10^{-8}$ & $5.41\times10^{-15}$ \\
\bottomrule
\end{tabular}
\end{table*}

\begin{figure}[!t]
 \centering
 \includegraphics[width=3.35in]{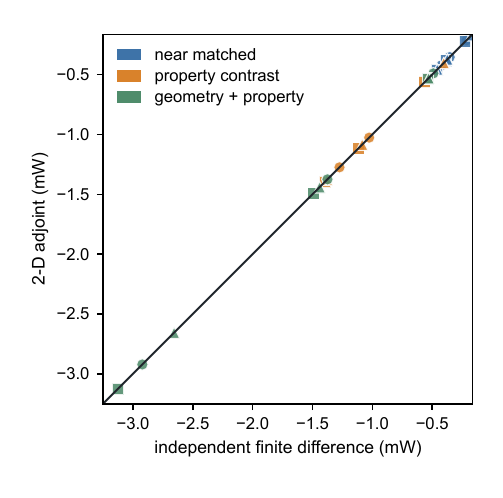}
 \caption{\label{fig:s_2d_verification}\textbf{Adjoint--finite-difference
 consistency.}  Two-dimensional adjoint responses are compared with
 independently evaluated centered finite differences of the same finite-volume
 residual for the 27 combinations of three mismatch families, three
 $\Gamma_m(T)$ profiles, and three sidewall couplings.  The diagonal is
 equality; colors distinguish the mismatch families.  The maximum relative
 difference is $3.96\times10^{-8}$.  Because both derivatives act on the same
 finite-volume constitutive model, this comparison tests the discrete
 linearization and adjoint, not the material laws.}
\end{figure}

Figure~\ref{fig:s_2d_verification} displays the case-by-case derivative comparison
under the same numerical conditions.
\clearpage

Grid refinement used the property-contrast family at unit sidewall coupling.
The medium-to-fine changes in the adjoint response were 0.1038\%, 0.0918\%,
and 0.1183\% for the constant, linear, and localized bases.  Adjoint and
finite-difference values remained much closer to each other on every fixed
grid (Table~\ref{tab:s_2d_grid}), showing that the residual grid dependence is
a spatial-discretization effect shared by both derivative routes rather than
an adjoint inconsistency.

\clearpage
\begin{table*}[!t]
\caption{Two-dimensional grid refinement for all three common-mode bases.}
\label{tab:s_2d_grid}
\small
\begin{tabular}{lcccc}
\toprule
\sftcell{0.18\textwidth}{$\Gamma_m$ basis and grid} & Adjoint (mW) &
\sftcell{0.16\textwidth}{Finite difference (mW)} &
\sftcell{0.16\textwidth}{Relative difference} &
\sftcell{0.16\textwidth}{Energy residual (W)} \\
\midrule
\sftcell{0.18\textwidth}{Constant, $16\times10$} & $-1.0231668$ & $-1.0231668$ & $1.04\times10^{-9}$ & $2.49\times10^{-15}$ \\
\sftcell{0.18\textwidth}{Constant, $32\times20$} & $-1.0277911$ & $-1.0277911$ & $1.24\times10^{-9}$ & $2.97\times10^{-15}$ \\
\sftcell{0.18\textwidth}{Constant, $64\times40$} & $-1.0288591$ & $-1.0288591$ & $1.24\times10^{-9}$ & $5.54\times10^{-15}$ \\
\sftcell{0.18\textwidth}{Linear, $16\times10$} & $-1.1127305$ & $-1.1127305$ & $1.39\times10^{-9}$ & $2.49\times10^{-15}$ \\
\sftcell{0.18\textwidth}{Linear, $32\times20$} & $-1.1172249$ & $-1.1172249$ & $1.58\times10^{-9}$ & $2.97\times10^{-15}$ \\
\sftcell{0.18\textwidth}{Linear, $64\times40$} & $-1.1182514$ & $-1.1182514$ & $1.64\times10^{-9}$ & $6.34\times10^{-15}$ \\
\sftcell{0.18\textwidth}{Localized, $16\times10$} & $-1.0883472$ & $-1.0883472$ & $1.95\times10^{-9}$ & $2.49\times10^{-15}$ \\
\sftcell{0.18\textwidth}{Localized, $32\times20$} & $-1.0835641$ & $-1.0835641$ & $2.41\times10^{-9}$ & $2.97\times10^{-15}$ \\
\sftcell{0.18\textwidth}{Localized, $64\times40$} & $-1.0822842$ & $-1.0822842$ & $2.12\times10^{-9}$ & $6.22\times10^{-15}$ \\
\bottomrule
\end{tabular}
\end{table*}

The finest representative p leg had a maximum transverse-to-axial
current-density ratio of 0.187 and a maximum lateral temperature span of
\SI{3.97}{\kelvin}; the local $\rho$ and $k$ fields spanned factors of 2.45
and 2.05, confirming substantial two-dimensional variation.

\subsection{Nonmonotonic field and topology controls}

A high-Joule control used the geometry--property-contrast family,
$|I|=\SI{3.0}{\ampere}$, equal \SI{300}{\kelvin} end temperatures, sidewall
multiplier 0.25, the linear $\Gamma_m$ basis, and a $32\times20$ grid.  The p
and n fields reached internal maxima of 371.19 and \SI{361.70}{\kelvin},
respectively, and each mean axial profile changed gradient sign once.  The
oriented-measure adjoint predicted $4.824$ mW per $\epsilon$, compared
with $4.824$ mW from the nonlinear finite difference, a relative
difference of $1.08\times10^{-8}$.  A single-valued $x(T)$ kernel would be
undefined for this control, whereas Eq.~(\ref{eq:s_measure}) remains valid.

\newpage
A shared-isothermal-pad constant shift of
$C=\SI{80}{\micro\volt\per\kelvin}$ changed aggregate $Q_c$, $Q_h$, side
heat, voltage, and electrical power only at the $10^{-16}$ SI-unit scale.  In
the same heterogeneous model, splitting the p and n cold pads by
\SI{4}{\kelvin} at $I=\SI{1.1}{\ampere}$ produced
$\Delta Q_c=\SI{0.352}{\milli\watt}$ and
$\Delta V_1=-\SI{0.320}{\milli\volt}$, with errors of
$6.75\times10^{-17}$ W and $9.80\times10^{-17}$ V relative to the exact
topology law derived below.  These controls connect the multidimensional
calculation to both the global null and the endpoint term.

The two-dimensional model resolves heterogeneous current flow, transverse heat
transport, sidewall loss, and nonmonotonic temperature fields with fixed-temperature
end pads.  Finite end-contact conductance, temperature-dependent $\rho$ and $k$,
radiation, contact mechanics, and three-dimensional package spreading require the
corresponding extensions.  In particular, a temperature-dependent $\rho$ would
couple the electrical field variation into the thermal tangent, and a
temperature-dependent $k$ would add conductivity-variation terms.  Neither
feedback is represented by the 27-case sweep above.  The following independent
calculation adds both effects while retaining the same steady scalar scope.

\subsection{Fully coupled temperature-dependent transport check}

We next solved a synthetic heterogeneous two-dimensional pair with
temperature-dependent electrical and thermal transport.  The geometry,
reference heterogeneity fields, sidewall conditions, signed current, and
linear common-mode basis are those of the property-contrast family at unit
sidewall multiplier in Table~\ref{tab:s_2d_design} and
Eq.~(\ref{eq:s_2d_bases}), with $T_c=300$ K, $T_h=350$ K, and
$|I|=1.1$ A; only the constitutive temperature dependence is changed:
\begin{equation}
 \begin{aligned}
 \rho(T,x,y)&=\rho_{\mathrm{ref}}(x,y)
 \exp[\SI{0.0032}{\per\kelvin}(T-\SI{325}{\kelvin})],\\
 k(T,x,y)&=k_{\mathrm{ref}}(x,y)
 \exp[-\SI{0.0024}{\per\kelvin}(T-\SI{325}{\kelvin})].
 \end{aligned}
 \label{eq:s_2d_variable_properties}
\end{equation}
Over 50 K these laws change $\rho$ by $+17.4\%$ and $k$ by $-11.3\%$.
They provide a moderate constitutive test and are not fitted representations
of either published material pair.

At every nonlinear residual evaluation, the electrical equation with
$\sigma(T,x,y)=1/\rho(T,x,y)$ was solved at fixed total current.  The resulting
divergence-free current field, facewise Joule heat, $k(T)$ conduction operator,
and conservative Thomson flux were then rebuilt.  Differentiating this reduced
residual after the electrical solution gives the complete electrothermal Schur
complement, including $\mathrm d\bm J/\mathrm dT$ and its Joule contribution;
the current field is not frozen in the thermal Jacobian.

\begin{table}[htbp]
\caption{Grid check for the fully coupled temperature-dependent
two-dimensional model.  The derivative is the paired cold-port response per
unit common-mode perturbation; nonlinear differences use
$\epsilon=\pm0.002$.}
\label{tab:s_2d_fully_coupled}
\small
\begin{tabular}{lccc}
\toprule
Grid per branch & Adjoint (mW) & Relative difference & Relative energy residual \\
\midrule
$8\times5$ & $-1.226046$ & $3.03\times10^{-8}$ & $4.66\times10^{-15}$ \\
$12\times7$ & $-1.239977$ & $2.95\times10^{-8}$ & $4.83\times10^{-14}$ \\
$16\times9$ & $-1.244191$ & $1.51\times10^{-8}$ & $8.21\times10^{-13}$ \\
\bottomrule
\end{tabular}
\end{table}

The medium-to-fine derivative change is 0.3387\%.  Across the three grids, the
largest adjoint--nonlinear finite-difference relative difference is
$3.03\times10^{-8}$.  Across the p- and n-branch systems, the discrete 2-norm
condition number ranges from 46.3 to 185.2, equivalently
$\sigma_{\min}/\sigma_{\max}\geq5.40\times10^{-3}$.  These are discrete
conditioning diagnostics rather than a continuum spectral margin.  When
$\rho_T$ is set exactly to zero in a matched control, the current-redistribution
contribution returns to numerical zero.
The branch-collection response therefore
persists when $\sigma(T)$ redistributes current and $k(T)$ changes the
conduction operator.  This synthetic calculation confirms the numerical
consistency of the fully coupled formulation; it does not validate a specific
material or package.

\section{Cr-doped PbSe case study from published transport data}

We digitized the p0.001 and n0.005 PbSe/Cr transport curves reported by Liu
\textit{et al.}\cite{liu2026pbse}  Seebeck coefficient and electrical conductivity
were extracted from vector graphics in the source article, while total thermal
conductivity was digitized from the Supplementary Information.  An independent
vector-graphics digitization provided a comparison of the Seebeck curves.  The property
roles and temperature ranges are summarized in Table~\ref{tab:s_pbse_sources}.
The resulting model represents the published curves over their common temperature
range rather than a calibrated sample-specific module.

Refs.~\onlinecite{liu2026pbse} and \onlinecite{liu2025bts} arose from the broader
Li-Dong Zhao collaboration, which includes contributors to the present work.
All calculations use publicly available figures, tables, and methods; no
unpublished measurements or calibration files enter the analysis.

The article describes the module as all-PbSe, whereas the Supplementary
Methods (printed p.~S7) use the phrase ``7-pair SnSe/BTS-based.''  Because the
public record does not provide a sample-to-leg mapping that resolves these two
descriptions, the seven-pair long-leg condition is used only as a nominal
geometry and operating scale.

Here PCHIP denotes a shape-preserving piecewise-cubic Hermite interpolant.

\begin{table*}[htbp]
\caption{PbSe/Cr input roles and temperature supports.}
\label{tab:s_pbse_sources}
\small
\begin{tabular}{lllll}
\toprule
\sftcell{0.14\textwidth}{Quantity} & \sftcell{0.14\textwidth}{Published figure} &
\sftcell{0.17\textwidth}{Temperature range} & \sftcell{0.18\textwidth}{Representation} &
\sftcell{0.21\textwidth}{Uncertainty treatment} \\
\midrule
\sftcell{0.14\textwidth}{$S_p,S_n$} & \sftcell{0.14\textwidth}{Article Fig. 1 vector markers} &
\sftcell{0.17\textwidth}{Seven points, approximately 300--573 K} &
\sftcell{0.18\textwidth}{PCHIP with analytic derivative; independent vector-graphics digitization for comparison} &
\sftcell{0.21\textwidth}{Reported ZEM-3 error kept below 5\%; not treated as a standard deviation} \\
\sftcell{0.14\textwidth}{$\sigma_p,\sigma_n$} & \sftcell{0.14\textwidth}{Article Fig. 1 vector markers} &
\sftcell{0.17\textwidth}{Seven points, approximately 300--573 K} &
\sftcell{0.18\textwidth}{PCHIP for $\rho=1/\sigma$} &
\sftcell{0.21\textwidth}{Reported error kept below 5\%; separated from Seebeck perturbations} \\
\sftcell{0.14\textwidth}{$\kappa_p,\kappa_n$} & \sftcell{0.14\textwidth}{Ref.~\onlinecite{liu2026pbse}, Fig. S9 (SI)} &
\sftcell{0.17\textwidth}{Seven points, approximately 300--573 K} &
\sftcell{0.18\textwidth}{PCHIP} &
\sftcell{0.21\textwidth}{Reported uncertainty within 15\%; not pooled with digitization} \\
\sftcell{0.14\textwidth}{Operating point} & \sftcell{0.14\textwidth}{Article Fig. 4a endpoint} &
\sftcell{0.17\textwidth}{$T_c\simeq310$ K, $T_h\simeq363$ K} &
\sftcell{0.18\textwidth}{Seven 2-by-2-by-6 mm$^3$ p/n pairs; $I\simeq2.81$ A} &
\sftcell{0.21\textwidth}{Entire field kept inside the common published range; unrounded coordinates retained in the repository} \\
\bottomrule
\end{tabular}
\end{table*}

The p/n curves were represented by PCHIP, with extrapolation disabled.  The analytic derivative of
the same interpolant supplied $T\,\mathrm dS/\mathrm dT$; no separately fitted
Thomson curve was introduced.  The p and n legs were each
$2\times2\times6$ mm$^3$ and were repeated for seven series pairs.  A
contact-resistance sensitivity used the reported p- and n-side specific
values, 26 and \SI{3}{\micro\ohm\centi\meter\squared}, with two interfaces per
leg and equal cold/hot Joule partition.  This gives a seven-pair series
resistance of \SI{0.0102}{\ohm}.  This value defines a contact-resistance
sensitivity case.

For a source curve pair, define
\begin{equation}
 M_0(T)=\frac{S_p(T)+S_n(T)}{2},\qquad
 \alpha_0(T)=S_p(T)-S_n(T).
 \label{eq:s_pbse_coordinates}
\end{equation}
The common-mode-flattened counterfactual is
\begin{equation}
 S_p^{\mathrm{flat}}(T)=M_0(T_c)+\frac{\alpha_0(T)}{2},\qquad
 S_n^{\mathrm{flat}}(T)=M_0(T_c)-\frac{\alpha_0(T)}{2}.
 \label{eq:s_pbse_flat}
\end{equation}
It preserves $\alpha_0(T)$ pointwise and leaves $\rho(T)$, $\kappa(T)$,
geometry, contacts, and reservoir temperatures unchanged.  The path from the
flattened baseline to the source curve is
\begin{equation}
 S_i^\epsilon(T)=S_i^{\mathrm{flat}}(T)
 +\epsilon\left[M_0(T)-M_0(T_c)\right],
 \label{eq:s_pbse_path}
\end{equation}
so the direct cold endpoint term is zero and $\epsilon=1$ restores the source
curves.  Over the operating interval, $M_0$ increases from 2.60 to
\SI{6.39}{\micro\volt\per\kelvin}, while the PCHIP-derived
$\Gamma_{\mathrm{cm}}$ spans 20.8--\SI{25.9}{\micro\volt\per\kelvin}.

\subsection{First-order, fixed-current finite, and reoptimized observables}

The three quantities in Table~\ref{tab:s_pbse_observables} answer different
questions.  The first-order value is
$\mathrm dQ_c/\mathrm d\epsilon|_{\epsilon=0}$ at fixed current and fixed
endpoints.  The finite value is $Q_c(\epsilon=1)-Q_c(0)$ at that same current.
The reoptimized value compares the independently optimized source and
flattened models.  Only the first is the theorem's local kernel.

\begin{table*}[htbp]
\caption{Separation of the three nominal PbSe/Cr common-mode observables.}
\label{tab:s_pbse_observables}
\small
\begin{tabular}{lll}
\toprule
\sftcell{0.22\textwidth}{Observable} & \sftcell{0.28\textwidth}{Response components (mW)} & \sftcell{0.34\textwidth}{Meaning} \\
\midrule
\sftcell{0.22\textwidth}{First order at $I=2.81$ A} & \sftcell{0.28\textwidth}{p: $+21.3$; n: $-17.9$; net: $+3.48$} & \sftcell{0.34\textwidth}{Derivative at the flattened baseline; fixed current and endpoints} \\
\sftcell{0.22\textwidth}{Finite $\epsilon:0\rightarrow1$ at the same current} & \sftcell{0.28\textwidth}{p: $+21.1$; n: $-18.0$; net: $+3.09$} & \sftcell{0.34\textwidth}{Nonlinear perturbation; 11.2\% below the first-order value} \\
\sftcell{0.22\textwidth}{Independent current reoptimization} & \sftcell{0.28\textwidth}{net: $+3.07$} & \sftcell{0.34\textwidth}{Difference of two maxima; source and flattened optima at 2.81 and 2.80 A} \\
\bottomrule
\end{tabular}
\end{table*}

The first-order absolute branch-transfer norm is 39.2 mW, so only 8.88\%
of the opposing p/n transfers survives at the aggregate cold port.  This
near-cancellation makes the residual especially sensitive to correlated
temperature-dependent errors.  A constant-property approximation at the same
operating point predicts 2.31 mW, or 66.3\% of the variable-property kernel,
and therefore does not capture the reconstructed collection measures.

\begin{figure*}[htbp]
 \centering
 \includegraphics[width=\textwidth]{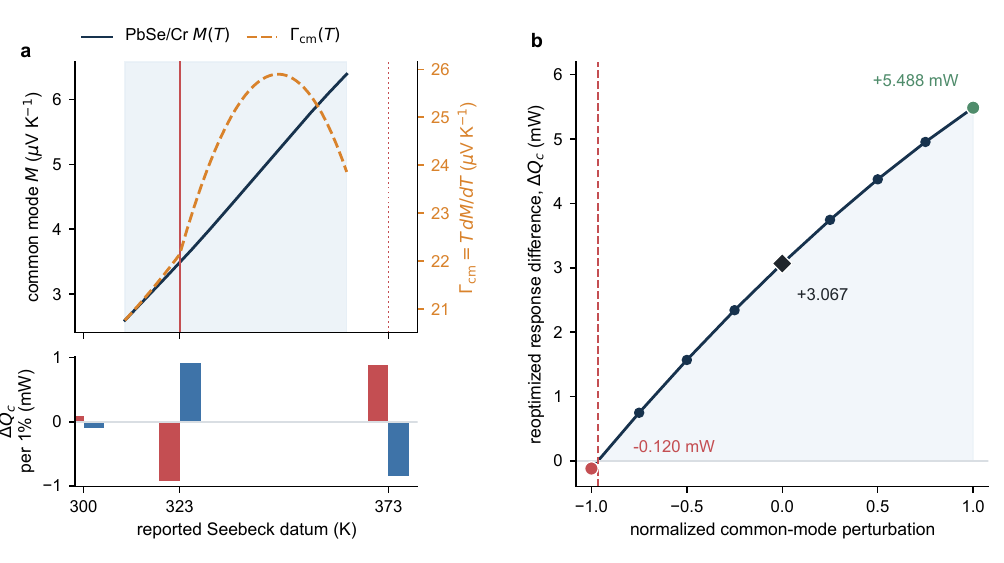}
 \caption{\label{fig:s_pbse_identifiability}\textbf{Published PbSe/Cr curves
 define a conditional common-mode identifiability test.} (a) The reconstructed
 material coordinates $M(T)$ and $\Gamma_{\mathrm{cm}}(T)$ over the reservoir
 interval, with local single-datum sensitivities below.  The sensitivity bars
 use signed 1\% relative changes of one p- or n-type datum; Table
 \ref{tab:s_pbse_knot_sensitivity} converts them to a common additive unit.
 (b) The finite, reoptimized source-minus-flat response along the
 $\alpha$-preserving perturbation path of
 Eq.~(\ref{eq:s_common_drift}).  The two smooth endpoint witnesses satisfy the
 reported marginal 5\% Seebeck bounds and give $-0.120$ and $+5.49$ mW.  The
 fixed-current first-order and finite fixed-current observables are reported
 separately in Table~\ref{tab:s_pbse_observables} and are not overlaid on this
 reoptimized curve.  The two endpoints delimit one deterministic path, not a
 confidence interval or a device-validation band.}
\end{figure*}

\section{Conditional sign identification under published PbSe/Cr marginal bounds}

The published Supplementary Information states that Seebeck and electrical-
conductivity measurement errors were kept below 5\% and that total thermal-
conductivity uncertainty was within 15\%.\cite{liu2026pbse}  These limits do
not specify probability distributions, p/n covariance, temperature covariance,
or the share attributable to common calibration drift.  We therefore use them
only as marginal deterministic bounds.  The calculation below is a
partial-identification test: it asks whether two admissible Seebeck curve pairs
can produce opposite response signs.  It is not a probability model, an error
bar, or a claim that the material itself changes $S$ while $\sigma$ and
$\kappa$ remain fixed.  Holding the other reconstructed inputs fixed isolates
the information supplied by the published Seebeck curves and their stated
marginal bounds.

The smooth common-mode perturbation changes both Seebeck laws by the same additive
drift
\begin{equation}
 \widetilde S_p(T)=S_p(T)+a z(T),\qquad
 \widetilde S_n(T)=S_n(T)+a z(T),\qquad
 z(T)=\frac{2T-T_{\min}-T_{\max}}{T_{\max}-T_{\min}}.
 \label{eq:s_common_drift}
\end{equation}
Thus $\widetilde\alpha(T)=\alpha(T)$ exactly, while
$\widetilde M(T)=M(T)+az(T)$.  Here $T_{\min}$ and $T_{\max}$ are the common
Seebeck support endpoints, so $z$ runs from $-1$ to $+1$ without assigning
spurious precision to digitized temperatures.  The amplitude
$|a|=\SI{9.79}{\micro\volt\per\kelvin}$ is the largest common additive
endpoint amplitude on this path that remains within 5\% of both Seebeck
magnitudes over the continuous common range.  Both signs retain the observed
p- and n-curve signs and monotonicity.  They give finite, reoptimized
source-minus-flat responses of $-0.120$ and $+5.49$ mW.  The zero crossing is
at normalized amplitude $-0.966$, or
$a\simeq-\SI{9.46}{\micro\volt\per\kelvin}$.  The existence of opposite-sign
members of this admissible set shows that the response sign is not point
identified by the published marginal bounds alone; it does not assign either
member a physical probability.

\begin{table*}[htbp]
\caption{Sensitivity of the PbSe/Cr response to common-mode Seebeck perturbations,
interpolation, digitization, and transport properties.}
\label{tab:s_pbse_witnesses}
\small
\begin{tabular}{lcl}
\toprule
\sftcell{0.25\textwidth}{Perturbation or model} & \sftcell{0.20\textwidth}{Reoptimized finite response difference (mW)} & \sftcell{0.43\textwidth}{Interpretation} \\
\midrule
\sftcell{0.25\textwidth}{Nominal PCHIP} & $+3.07$ & \sftcell{0.43\textwidth}{Central values extracted from the published figure} \\
\sftcell{0.25\textwidth}{Smooth common-mode perturbation, $a=-9.79~\mu$V K$^{-1}$} & $-0.120$ & \sftcell{0.43\textwidth}{Pointwise $\alpha$ preserved; signs, monotonicity, and the continuous marginal 5\% bound retained} \\
\sftcell{0.25\textwidth}{Smooth common-mode perturbation, $a=+9.79~\mu$V K$^{-1}$} & $+5.49$ & \sftcell{0.43\textwidth}{Positive-shift member of the same deterministic path} \\
\sftcell{0.25\textwidth}{Piecewise-linear interpolation} & $+3.10$ & \sftcell{0.43\textwidth}{Same seven published markers} \\
\sftcell{0.25\textwidth}{Smooth quadratic within the measurement bound} & $+2.80$ & \sftcell{0.43\textwidth}{Marker residuals below 5\%; carrier signs and monotonicity retained} \\
\sftcell{0.25\textwidth}{Independent vector-graphics digitization} & $+3.05$ & \sftcell{0.43\textwidth}{Extraction-route repeatability; not a measurement-accuracy estimate} \\
\sftcell{0.25\textwidth}{Monotonic locally active PCHIP combinations} & $[-28.6,+10.4]$ & \sftcell{0.43\textwidth}{45 evaluated binary combinations; deterministic sensitivity range} \\
\sftcell{0.25\textwidth}{Separate $\sigma/\kappa$ combinations} & $[-0.958,+6.65]$ & \sftcell{0.43\textwidth}{$\sigma\pm5\%$, $\kappa\pm15\%$ with nominal Seebeck; evaluated separately from the Seebeck perturbations} \\
\bottomrule
\end{tabular}
\end{table*}

A common constant offset of the same scale changes the response by at most
$4.44\times10^{-16}$ W, confirming that the unresolved coordinate is the
temperature-dependent common-mode variation rather than an absolute constant alone.
The two extraction routes differ by at most 0.0077\% in n-type and 0.0051\%
in p-type Seebeck magnitude.  These values quantify repeatability between the
two vector-graphics extraction routes only; they are not estimates of marker
accuracy, experimental uncertainty, or sample reproducibility.  Within this
limited comparison, the plotted sign reversal is not caused by the extraction
route.

For the binary shape-preserving-interpolation analysis, only the first four of seven
reported data points can affect the
approximately 310--363 K target field.  Of the $2^8=256$ sign combinations, the
observed carrier-sign and monotonicity constraints retain nine valid p corners
and five valid n corners, or 45 evaluated pairs.  The extrema
in Table~\ref{tab:s_pbse_witnesses} are complete over these binary corners but
do not certify extrema over every continuous function within the reported
measurement bound.

Single-knot derivatives localize the missing information.  Table
\ref{tab:s_pbse_knot_sensitivity} reports the response to a positive additive
$+1~\mu$V K$^{-1}$ change of one datum, rather than to a signed relative
change.  The p and n columns can therefore be added for a common-mode knot
shift.  This
conversion matters because $S_n<0$: a $+1\%$ relative change makes the n-type
datum more negative, whereas a positive additive change makes it less
negative.  Numerically, the conversion is
$D_i^{\mathrm{add}}=D_i^{1\%}/(0.01S_i)$ when $S_i$ is expressed in
$\mu$V K$^{-1}$.  In the additive convention, the p- and n-datum sensitivities have
the same sign at both 323 K and 373 K.  A common p/n drift therefore reinforces
locally at a given temperature; the balance between the 323 and 373 K windows
is controlled by temperature covariance.  The reported 373 K datum lies just
above the hot reservoir temperature but influences the local
shape-preserving interpolation within the operating interval.

\begin{table}[htbp]
\caption{Fixed-current PbSe/Cr common-mode response to an additive
$+1~\mu$V K$^{-1}$ change at one Seebeck datum.  Data at or above 473 K lie
outside the active temperature interval and therefore have zero derivative in
this local shape-preserving interpolation.}
\label{tab:s_pbse_knot_sensitivity}
\begin{tabular}{ccc}
\toprule
Datum temperature (K) & p datum (mW per $\mu$V K$^{-1}$) & n datum (mW per $\mu$V K$^{-1}$) \\
\midrule
300 & $+0.044$ & $+0.050$ \\
323 & $-0.433$ & $-0.451$ \\
373 & $+0.373$ & $+0.379$ \\
423 & $+0.015$ & $+0.022$ \\
$\geq473$ & $0$ & $0$ \\
\bottomrule
\end{tabular}
\end{table}

Summing the p and n columns gives common-mode knot sensitivities of
$-0.884$ mW and $+0.752$ mW per $\mu$V K$^{-1}$ at 323 and 373 K,
respectively.  Their opposite signs across temperature explain why the
temperature covariance, rather than the marginal error of either curve alone,
controls the small terminal residual.

\subsection{Joint deterministic error geometry}

The marginal screens above do not specify how Seebeck, electrical-conductivity,
and thermal-conductivity errors can occur together.  We therefore constructed a
nine-dimensional positive-semidefinite ellipsoidal geometry with normalized
coordinate
\begin{equation}
 \bm x=(x_{\mathrm{inst}},x_{p,323},x_{n,323},x_{p,373},x_{n,373},
 x_{\sigma p},x_{\sigma n},x_{\kappa p},x_{\kappa n}).
 \label{eq:s_joint_coordinates}
\end{equation}
The shared coordinate produces the additive drift
\begin{equation}
 \delta S_{p,\mathrm{inst}}(T)=\delta S_{n,\mathrm{inst}}(T)
 =x_{\mathrm{inst}}a_{\max}z(T),\qquad
 a_{\max}=\SI{9.79}{\micro\volt\per\kelvin},
 \label{eq:s_joint_shared_drift}
\end{equation}
where $z(T)$ is defined in Eq.~(\ref{eq:s_common_drift}).  This coordinate
represents a shared thermovoltage or $\Delta T$ calibration-map systematic,
not a carrier-concentration or band-structure change.  It is therefore kept
independent of the $\sigma$ and $\kappa$ coordinates.  The public record does
not establish that the p- and n-type specimens were measured in one common
calibration state.

The four branch-specific Seebeck coordinates act on the dominant 323 and 373 K
knots with 5\% axis scales.  The remaining coordinates scale the complete p/n
$\sigma$ curves by 5\% and the complete p/n $\kappa$ curves by 15\%.  We write
\begin{equation}
 \bm x=\bm D(f)\bm C^{1/2}\bm u,\qquad \|\bm u\|_2\leq1,\qquad
 \bm D(f)=\operatorname{diag}
 \left(f,\sqrt{1-f^2}\,\bm{1}_4,\bm{1}_4\right).
 \label{eq:s_joint_ellipsoid}
\end{equation}
Here $\bm{1}_4$ denotes four repeated diagonal entries.  Thus $f$ assigns the
quadratic Seebeck budget between the shared additive drift and the four
branch-specific residuals.  Each matrix $\bm C$ is generated from
explicit factor loadings and is positive semidefinite.  Its entries define one
deterministic correlation-shape scenario; they are not empirical correlations,
probabilities, or a confidence region.  Every evaluated candidate is also
intersected with the continuous 5\% Seebeck ceiling and preserves the carrier
sign and monotonicity of both source curves.

For every candidate, the source and its pointwise $\alpha$-preserving flattened
counterfactual are reoptimized independently.  Table
\ref{tab:s_joint_error_geometry} reports extrema from a predefined set of perturbation directions
containing the local response-support directions and both signs of the
coordinate extremizers.  The intervals are deterministic nonlinear witnesses,
not certified extrema over the full ellipsoid.

\begin{table*}[htbp]
\caption{Joint deterministic PbSe/Cr error scenarios.  Each entry is the
evaluated interval of the reoptimized source-minus-flat response in mW.  The
shape triplet is $(\rho_{pn}^{S},\rho_{S\sigma},\rho_{\sigma\kappa})$; all
scenario matrices are positive semidefinite.}
\label{tab:s_joint_error_geometry}
\small
\begin{tabular}{lllll}
\toprule
\sftcell{0.17\textwidth}{Scenario} &
\sftcell{0.205\textwidth}{Correlation-shape triplet} &
\sftcell{0.155\textwidth}{$f=0$} &
\sftcell{0.155\textwidth}{$f=0.5$} &
\sftcell{0.155\textwidth}{$f=1$} \\
\midrule
\sftcell{0.17\textwidth}{Independent residual axes} &
\sftcell{0.205\textwidth}{$(0,0,0)$} &
\sftcell{0.155\textwidth}{$[-8.42,+10.2]$} &
\sftcell{0.155\textwidth}{$[-6.50,+10.0]$} &
\sftcell{0.155\textwidth}{$[-0.120,+7.45]$} \\
\sftcell{0.17\textwidth}{Same-run scale dominated} &
\sftcell{0.205\textwidth}{$(0.85,0,0.25)$} &
\sftcell{0.155\textwidth}{$[+0.334,+7.70]$} &
\sftcell{0.155\textwidth}{$[+0.340,+7.63]$} &
\sftcell{0.155\textwidth}{$[-0.120,+7.40]$} \\
\sftcell{0.17\textwidth}{Moderate band--transport coupling} &
\sftcell{0.205\textwidth}{$(0.60,-0.346,0.490)$} &
\sftcell{0.155\textwidth}{$[-0.198,+8.20]$} &
\sftcell{0.155\textwidth}{$[-0.056,+8.07]$} &
\sftcell{0.155\textwidth}{$[-0.120,+7.65]$} \\
\sftcell{0.17\textwidth}{Strong band-locked coupling} &
\sftcell{0.205\textwidth}{$(0.80,-0.335,0.671)$} &
\sftcell{0.155\textwidth}{$[+0.648,+6.98]$} &
\sftcell{0.155\textwidth}{$[+0.599,+7.24]$} &
\sftcell{0.155\textwidth}{$[-0.120,+7.83]$} \\
\bottomrule
\end{tabular}
\end{table*}

For the reoptimized observable used in this joint model, a positive additive
$1~\mu$V K$^{-1}$ change gives p/n contributions of $-0.450/-0.434$ mW at
323 K and $+0.408/+0.344$ mW at 373 K.  The branch contributions therefore
have the same sign at a given temperature; the partial cancellation is between
the 323 and 373 K windows, whose sums are $-0.884$ and $+0.752$ mW per
$\mu$V K$^{-1}$.  Along the pure negative shared-drift axis, the response
crosses zero at 0.9662 of the full 5\% axis, corresponding to an endpoint
amplitude of $-9.46~\mu$V K$^{-1}$.

The independent scenario contains negative witnesses for all three $f$ values.
The moderate scenario also contains one at $f=0.5$ ($-0.056$ mW), with
branch-specific Seebeck, $\sigma$, and $\kappa$ coordinates all nonzero.  In
contrast, the same-run and strong predefined direction sets retain positive minima
at $f=0$ and 0.5.  Those positive tested minima do not prove positivity over
the full nonlinear ellipsoid.  The public marginal statements therefore do not
select one sign conclusion: several explicit joint models cross zero, while
the tested strongly correlated libraries can narrow the response without
establishing global sign identification.

Resolving these alternatives requires paired same-run p/n thermovoltage and
$\Delta T$ records near 323 and 373 K, calibration and reference records,
replicate covariance, matched $\sigma/\kappa$ specimen identifiers, and a
traceable specimen-to-device mapping.  The analysis identifies the governing
temperature windows but does not estimate an empirical covariance matrix.

\section{Published-data requirements for pair-resolved comparison}

The PbSe/Cr Supplementary Information reports calculated cooling capacity and
coefficient of performance for 36 combinations of legs labelled N1--N6 and
P1--P6.\cite{liu2026pbse}  The public figures do not provide a complete
sample-preserving map from these labels to the $S(T)$, $\sigma(T)$, and
$\kappa(T)$ curves, nor the current and endpoint temperatures for every matrix
entry.  The rounded pair matrices therefore cannot be converted uniquely into
the branch collection measures defined in Eq.~(\ref{eq:s_measure}).  We do not
use them for a pair ranking or for a statistical correlation claim.

A branch-resolved comparison would require numerical transport tables tied to
each device leg, same-run p/n Seebeck repeats and reference calibration,
cross-leg and cross-temperature covariance, and the operating conditions for
each pair calculation.  The theorem therefore identifies the measurements needed
to distinguish a small terminal
residual from the much larger opposing branch transfers.

\section{Exact split-pad/reference-contrast topology law}

The endpoint term in Eq.~(\ref{eq:s_pair_theorem}) becomes directly
observable when corresponding p and n endpoints are maintained at different
temperatures and the semiconductor branches are changed relative to calibrated
fixed reference leads.  For element $j$, let the constant contrast be $C_j$,
the local p-branch current be $I_j$ (so the n-branch current is $-I_j$), and
define
\begin{equation}
 \Delta T_{c,j}=T_{c,p,j}-T_{c,n,j},\qquad
 \Delta T_{h,j}=T_{h,p,j}-T_{h,n,j},\qquad
 \Delta V_j\equiv V_{p,j}-V_{n,j}.
 \label{eq:s_split_definitions}
\end{equation}
Because $\Gamma_m=0$ for a constant contrast, the temperature fields are
unchanged at fixed endpoints and the boundary terms give
\begin{align}
 \boxed{\Delta Q_{c,\Sigma}}&=\boxed{\sum_jC_jI_j\Delta T_{c,j}},
 \label{eq:s_split_qc}\\
 \boxed{\Delta Q_{h,\Sigma}}&=\boxed{\sum_jC_jI_j\Delta T_{h,j}},
 \label{eq:s_split_qh}\\
 \boxed{\Delta P_{\mathrm{in},\Sigma}}&=
 \boxed{\sum_j I_j\Delta V_j}
 =\boxed{\sum_jC_jI_j(\Delta T_{h,j}-\Delta T_{c,j})}.
 \label{eq:s_split_power}
\end{align}
Thus the contrast-induced element voltage is
$\Delta V_j=C_j(\Delta T_{h,j}-\Delta T_{c,j})$ under the stated sign
convention.
Here $Q_{c,\Sigma}$ is the aggregate heat over two separately controlled,
generally nonisothermal cold-side node classes per element.  It is therefore
distinct from heat extracted from one isothermal cold reservoir.  The
increments satisfy the elementwise and aggregate energy identity
\begin{equation}
 \Delta Q_{h,\Sigma}-\Delta Q_{c,\Sigma}
 -\Delta P_{\mathrm{in},\Sigma}=0.
 \label{eq:s_split_energy}
\end{equation}
For a series array with common $C$ and $I$, define signed mean endpoint splits
$\overline{\Delta T_s}=N^{-1}\sum_j\Delta T_{s,j}$.  Then
\begin{align}
 \Delta Q_{c,\Sigma}&=CIN\overline{\Delta T_c},\nonumber\\
 \Delta Q_{h,\Sigma}&=CIN\overline{\Delta T_h},\nonumber\\
 \Delta V_\Sigma\equiv\sum_j\Delta V_j
 &=CN(\overline{\Delta T_h}-\overline{\Delta T_c}),\nonumber\\
 \Delta P_{\mathrm{in},\Sigma}&=I\Delta V_\Sigma.
 \label{eq:s_split_series}
\end{align}
For shared isothermal hot pads, $\overline{\Delta T_h}=0$, these equations give
the exact heat--voltage identity
\begin{equation}
 \boxed{I\Delta V_\Sigma=-\Delta Q_{c,\Sigma}.}
 \label{eq:s_split_interlock}
\end{equation}
For branched arrays with different $I_j$, the power sum in
Eq.~(\ref{eq:s_split_power}) is the general object; a single series terminal
voltage need not exist.

Three strict controls follow immediately.  The heat increments vanish at
$I_j=0$.  Shared corresponding endpoint temperatures give
$\Delta Q_{c,\Sigma}=\Delta Q_{h,\Sigma}=\Delta V_\Sigma=0$.  Finally, if the same
$C$ is also applied to every reference lead, interconnect, and external active
segment, the omitted paths supply the exact counterterm and recover the global
network null of Eq.~(\ref{eq:s_node_null}).  The split-pad law therefore does
not contradict the global co-shift theorem; it describes a calibrated
branch-to-reference contrast.

\subsection{Variable-property response and device scale}

The temperature-dependent one-dimensional model was evaluated for 36
heterogeneous cases spanning
$C\in\{-100,+80\}~\mu$V K$^{-1}$,
$I\in\{-1.3,0,+1.3\}$ A, and
$(\Delta T_c,\Delta T_h)=(0,0),(6,0),(-6,0),(0,4),(5,-3),(-5,3)$ K.
The p and n legs had different temperature-dependent $S$, $\rho$, and
$\kappa$.  The largest absolute errors relative to
Eqs.~(\ref{eq:s_split_qc})--(\ref{eq:s_split_power}) were
$7.16\times10^{-17}$ W in cold heat,
$5.55\times10^{-17}$ W in hot heat, and
$2.30\times10^{-17}$ V in voltage.  The maximum incremental energy residual
was $1.28\times10^{-16}$ W, and the maximum temperature-field change under the
constant contrast was $1.14\times10^{-13}$ K.

For scale only, the seven-pair PbSe/Cr nominal operating point was used as an
isothermal-module normalization.  It was not assumed that the reported device
had split pads or a calibrated $C$ contrast.  The special-case prediction and
crossings are summarized in Table~\ref{tab:s_topology_scale}.

\begin{table}[htbp]
\caption{Split-pad response scale at the nominal seven-pair operating point.}
\label{tab:s_topology_scale}
\small
\begin{tabular}{lr}
\toprule
\sftcell{0.58\textwidth}{Quantity} & Value \\
\midrule
\sftcell{0.58\textwidth}{Pair count and current} & $N=7$, $I=2.81$ A \\
\sftcell{0.58\textwidth}{Reference isothermal-module $Q_c$} & 0.664 W \\
\sftcell{0.58\textwidth}{Contrast and coherent signed mean split} & $C=80~\mu$V K$^{-1}$, $\overline{\Delta T_c}=5$ K \\
\sftcell{0.58\textwidth}{Predicted $\Delta Q_{c,\Sigma}$} & 7.87 mW \\
\sftcell{0.58\textwidth}{Fraction of reference $Q_c$} & 1.19\% \\
\sftcell{0.58\textwidth}{Predicted series $\Delta V_\Sigma$ for $\overline{\Delta T_h}=0$} & $-2.80$ mV \\
\sftcell{0.58\textwidth}{Split for a 1-mW model scale at the same $C,I,N$} & 0.635 K \\
\sftcell{0.58\textwidth}{Split for 1\% of reference $Q_c$ at the same $C,I,N$} & 4.22 K \\
\bottomrule
\end{tabular}
\end{table}

The 1-mW and 1\% lines are deterministic model scales, not calorimeter
detection limits or significance thresholds.  Define the signed coherence
factor as $\chi_c=\overline{\Delta T_c}/(5~\mathrm K)$ at this reference scale.
Incoherent pairwise splits cancel physically: $\chi_c=1$, 0.5, 0.25, and 0 give 7.87, 3.94,
1.97, and 0 mW at the reference scenario.

For a 10\%-precision test at the representative point, let $u_Q$ and $u_V$
denote independent standard uncertainties of the differential heat and voltage
measurements.  The paired heat--voltage residual then requires
$[u_Q^2+(Iu_V)^2]^{1/2}<0.79$ mW.  Equal uncertainty allocation corresponds to
$u_Q<0.56$ mW and $u_V<0.20$ mV at $I=2.81$ A; the heat-flow and endpoint-
stability channels are therefore the more demanding parts of the test.

\subsection{Experimental protocol and current parity}

An exact contrast test requires two calibrated reference-lead states whose
thermopower difference is $-C$.  Lead resistance, heat leakage, contact state,
and endpoint temperatures must be matched or corrected.  A suitable sequence
is to stabilize a shared hot block, impose both signs of the p/n cold split,
and measure the local cold- and hot-side heat rates and the series four-terminal
voltage $\Delta V_\Sigma$ at $+I$ and $-I$.  The calibrated reference state
should then be reversed, followed by a repeat measurement at zero split.  For
the same-current series case, simultaneous verification of
$\Delta Q_{h,\Sigma}-\Delta Q_{c,\Sigma}=I\Delta V_\Sigma$ is stronger than a cold-heat observation
alone; for isothermal hot pads it reduces to
Eq.~(\ref{eq:s_split_interlock}).

The current-odd projection removes only components of the measured
reference-state contrast that are even in current.  It does not automatically
remove baseline Peltier heat, Thomson redistribution, contact Peltier heat, or
odd thermal feedback.  Those contributions must be bounded by calibrated
state subtraction and independent resistance, thermal-leak, contact, and
temperature measurements.  Reversing the pad split on one unchanged device is
also not an exact $C$ switch because it changes conduction and the
self-consistent temperature field.  The combined heat--voltage relation
therefore provides the stronger experimental test; split-pad heat alone is not
unique to common-mode thermopower.

\section{Bi$_2$Te$_3$-based cross-material application}

To examine the transfer law in a second chemistry, we constructed a model from
the source-labelled n-type BTS+0.2\%Cu (Bi$_2$Te$_3$-based) and commercial
p-type BST (Bi--Sb--Te) reported by Liu \textit{et al.}\cite{liu2025bts}.
The n-branch $S$, $\sigma$, and total $\kappa$ curves were reconstructed from
the article figures, and the corresponding p-branch curves from Supplementary
Fig. S9 of Ref.~\onlinecite{liu2025bts}.
Each property has six published points over 300--523 K.  The calculations used the
published seven-pair leg geometry, fixed material endpoints
$(T_c,T_h)=(323,373)$ K, and $I=3$ A.  The model did not fit unknown thermal
boundary parameters or extrapolate the published transport curves below 300 K.
All fields over 0--3 A remained within 323--373 K, and
the normalized perturbation remained between its endpoint zeros and unit
interior maximum.

The discriminating perturbation was the endpoint-zero common mode
\begin{equation}
 m_{\mathrm{test}}(T)=4M_{\mathrm{pk}}
 \frac{(T-T_c)(T_h-T)}{(T_h-T_c)^2},
 \qquad M_{\mathrm{pk}}=\SI{10}{\micro\volt\per\kelvin},
 \label{eq:s_endpoint_zero_test}
\end{equation}
added equally to p and n.  This construction preserves $\alpha(T)$ pointwise,
leaves the open-circuit voltage unchanged, and removes both endpoint Peltier
terms.  Numerically, the maximum pointwise change in $\alpha$ was
$5.42\times10^{-20}$ V K$^{-1}$, the open-circuit-voltage change was
$9.49\times10^{-21}$ V in magnitude, and the mode was zero at both material
endpoints.  The remaining response therefore probes the distributed
$\Gamma_m$ action on the difference of the p- and n-branch collection measures.

\begin{table}[htbp]
\caption{BTS/BST endpoint-zero analysis at fixed material endpoints.}
\label{tab:s_bts_bst}
\small
\begin{tabular}{ll}
\toprule
\sftcell{0.38\textwidth}{Quantity} & \sftcell{0.48\textwidth}{Result} \\
\midrule
\sftcell{0.38\textwidth}{Material and field window} & \sftcell{0.48\textwidth}{n-BTS+0.2\%Cu/p-BST; $323\leq T\leq373$ K inside the published 300--523 K range} \\
\sftcell{0.38\textwidth}{Endpoint-zero perturbation} & \sftcell{0.48\textwidth}{$M_{\mathrm{pk}}=10~\mu$V K$^{-1}$, with exact cold/hot zeros and pointwise $\delta\alpha=0$} \\
\sftcell{0.38\textwidth}{p/n adjoint branch transfers per pair} & \sftcell{0.48\textwidth}{$+5.44$ and $-5.26$ mW} \\
\sftcell{0.38\textwidth}{Nominal baseline module $Q_c$} & \sftcell{0.48\textwidth}{$1.34$ W} \\
\sftcell{0.38\textwidth}{Nominal aggregate response at 3 A} & \sftcell{0.48\textwidth}{$+1.31$ mW by the adjoint and by an independent central difference} \\
\sftcell{0.38\textwidth}{Adjoint--finite-difference relative difference} & \sftcell{0.48\textwidth}{$2.33\times10^{-7}$; maximum relative energy residual $2.47\times10^{-16}$} \\
\sftcell{0.38\textwidth}{Exact negative controls} & \sftcell{0.48\textwidth}{Zero current, a global constant co-shift, and identical-transport antisymmetric branches each give zero response} \\
\sftcell{0.38\textwidth}{Six-coordinate combinations at reported marginal bounds} & \sftcell{0.48\textwidth}{64 deterministic combinations from branch-level $S/\sigma/\kappa$ gains at $5/5/8\%$: $[-2.91,+5.73]$ mW} \\
\bottomrule
\end{tabular}
\end{table}

The nominal per-pair branch transfers oppose one another so strongly that only
1.75\% of their absolute sum survives in the aggregate response.  A separate
nonlinear boundary-value solution verifies the adjoint derivative for this
BTS/BST model.  Independent enumeration of the reported marginal 5\%, 5\%,
and 8\% bounds for $S$, $\sigma$, and $\kappa$ spans $-2.91$ to $+5.73$ mW.
Because the source does not specify their joint covariance, this deterministic
set shows conditional non-identification of the sign; it is not a probability
interval.
Because the reported below-300-K device operation lies outside the available
transport range, this result is restricted to the 323--373 K materials model.

\section{Numerical methods and data availability}

The temperature-dependent one-dimensional legs were solved as conservative
boundary-value problems in total heat and temperature using adaptive
collocation.  Tabulated properties were evaluated only inside their specified
closed supports; PCHIP extrapolation was disabled.  The adjoint equation was
integrated independently as a two-point shooting problem.  Central differences
used independent nonlinear solutions at both perturbation signs.  The finite-contact
calculation solved the two reservoir node balances and the bulk legs
self-consistently.  The two-dimensional formulation used sparse
finite-volume matrices, conservative edge Joule power, flux-form Thomson
transport, Newton iteration, and the transpose of the complete tangent
operator.  In the fully coupled temperature-dependent case, the electrical
field and Joule power were recomputed at every thermal residual evaluation;
the reduced Jacobian was differentiated only after this fixed-current
electrical solve, thereby retaining $\sigma_T$-driven current redistribution
and the $k_T$ conduction terms.

Quantities reconstructed from published figures and the resulting physical
response scales are displayed with three or four significant figures.  Extra
digits are retained only in the numerical-verification tables when needed to
resolve convergence or derivative differences; unrounded machine-readable
values are reserved for the public data package.

The final versioned package of processed data, numerical outputs, and scripts
supporting this work will be released with the version of record when the
article is published in a peer-reviewed journal. Materials required for
editorial and referee assessment are available from the corresponding authors
during review. Publisher-hosted source articles used for figure digitization
remain available through the cited references and are not redistributed. The
reported ranges are deterministic numerical results rather than statistical
confidence intervals.

\section{Scope of the physical conclusions}

Within steady scalar local-equilibrium transport, the exact global co-shift
null and the first-order branch-relative theorem separate the baseline material
coordinate from its test direction.  Four quantities organize the response:
$\alpha(T)$ supplies differential thermoelectric drive;
$\Gamma_{\mathrm{cm}}(T)$ describes the material common mode, with
$\delta\Gamma_{\mathrm{cm}}=\Gamma_m$ for the branch-relative perturbation;
$\mu_p-\mu_n$ determines its collection at a port; and endpoint topology
controls the direct term.  The closed-form, finite-contact, spatially
heterogeneous fixed-$\rho/\kappa$, and fully coupled temperature-dependent
two-dimensional calculations examine complementary limits of this statement.
The PbSe/Cr application shows that the nominal port response is a small
difference of large opposing transfers.  Several explicit positive-semidefinite
joint-error geometries contain sign-reversing witnesses, whereas the tested
same-run and strongly correlated predefined direction sets retain positive minima at
smaller shared-drift allocations.  The latter are not global positivity proofs.
The published marginal bounds therefore do not select a unique sign, but the
calculation does not claim sign indeterminacy for every possible joint model.
The BTS/BST calculation applies the same endpoint-zero perturbation to a second
material family and shows similar conditional sensitivity to the published
property bounds.

The results establish a paired-branch cancellation condition and a
topology-resolved heat--voltage relation for steady scalar transport.
Applications to PbSe/Cr and BTS/BST show both the transfer scale and the need
for sample-matched Seebeck data.  Extension to magnetic tensors, transient
storage, active shunts, nonideal contact thermopower, or full three-dimensional
packaging requires the corresponding transport terms.

\clearpage

\bibliographystyle{aipnum4-2}
\bibliography{references}